\documentclass[longauth]{aa}
\usepackage{graphicx}
\usepackage{txfonts}
\usepackage{subfigure}
\usepackage{array}
\usepackage{float}
\usepackage{flafter}
\usepackage{stfloats}
\usepackage{afterpage}
\usepackage{multicol}
\usepackage[section]{placeins}

\usepackage[colorlinks=true,urlcolor=blue,linkcolor=blue,citecolor=blue]{hyperref}

\begin{document}

\title{A large sub-Neptune transiting the thick-disk M4~V TOI-2406}

\subtitle{}

\author{
    % 1
    R.D.~Wells\inst{\ref{unibe},\thanks{robert.wells@csh.unibe.ch}}
    \and B.V.~Rackham\inst{\ref{miteaps},\ref{mitkavli},\thanks{51 Pegasi b Fellow}}
    \and N.~Schanche\inst{\ref{unibe}}
    \and R.~Petrucci\inst{\ref{cordoba},\ref{conicet}}
    \and Y.~G\'omez~Maqueo~Chew\inst{\ref{ciudad}}
    \and B.-O.~Demory\inst{\ref{unibe}}
    \and A.J.~Burgasser\inst{\ref{ucsd}}
    \and R.~Burn\inst{\ref{mpia}}
    \and F.J.~Pozuelos\inst{\ref{aru_liege},\ref{star_liege}}
    \and M.N.~G\"unther\inst{\ref{mitkavli},\ref{mitphysics},\thanks{Juan Carlos Torres Fellow}}
    \and L.~Sabin\inst{\ref{uname}}
    \and U.~Schroffenegger\inst{\ref{unibe}}
    \and M.A.~G\'omez-Mu\~noz\inst{\ref{uname}}
    \and K.G.~Stassun\inst{\ref{vanderbilt}}
    \and V.~Van~Grootel\inst{\ref{star_liege}}
    \and S.B.~Howell\inst{\ref{ames}}
    \and D.~Sebastian\inst{\ref{birmingham}}
    \and A.H.M.J.~Triaud\inst{\ref{birmingham}}
    \and D.~Apai\inst{\ref{arizona},\ref{nexus},\ref{lpl}}
    % 2
    \and I.~Plauchu-Frayn\inst{\ref{uname}}
    \and C.A.~Guerrero\inst{\ref{uname}}
    \and P.F.~Guill\'en\inst{\ref{uname}}
    \and A.~Landa\inst{\ref{uname}}
    \and G.~Melgoza\inst{\ref{uname}}
    \and F.~Montalvo\inst{\ref{uname}}
    \and H.~Serrano\inst{\ref{uname}}
    \and H.~Riesgo\inst{\ref{uname}}
    % 3
    \and K.~Barkaoui\inst{\ref{aru_liege},\ref{oukaimeden}}
    \and A.~Bixel\inst{\ref{arizona}}
    \and A.~Burdanov\inst{\ref{miteaps}}
    \and W.P.~Chen\inst{\ref{ncu_taiwan}}
    \and P.~Chinchilla\inst{\ref{aru_liege},\ref{iac}}
    \and K.A.~Collins\inst{\ref{cfa}}
    \and T.~Daylan\inst{\ref{mitkavli},\ref{mitphysics},\thanks{Kavli Fellow}}
    \and J.~de~Wit\inst{\ref{miteaps}}
    \and L.~Delrez\inst{\ref{aru_liege},\ref{star_liege}}
    \and M.~D\'evora-Pajares\inst{\ref{granada}}
    \and J.~Dietrich\inst{\ref{arizona}}
    \and G.~Dransfield\inst{\ref{birmingham}}
    \and E.~Ducrot\inst{\ref{aru_liege}}
    \and M.~Fausnaugh\inst{\ref{mitkavli},\ref{mitphysics}}
    \and E.~Furlan\inst{\ref{nesi}}
    \and P.~Gabor\inst{\ref{vat_arizona}}
    \and T.~Gan\inst{\ref{tsinghua}}
    \and L.~Garcia\inst{\ref{aru_liege}}
    \and M.~Ghachoui\inst{\ref{oukaimeden}}
    \and S.~Giacalone\inst{\ref{berkeley}}
    \and A.B.~Gibbs\inst{\ref{cali}}
    \and M.~Gillon\inst{\ref{aru_liege}}
    \and C.~Gnilka\inst{\ref{ames}}
    \and R.~Gore\inst{\ref{berkeley}}
    \and N.~Guerrero\inst{\ref{mitkavli},\ref{mitphysics}}
    \and T.~Henning\inst{\ref{mpia}}
    \and K.~Hesse\inst{\ref{mitkavli},\ref{mitphysics}}
    \and E.~Jehin\inst{\ref{star_liege}}
    \and J.M.~Jenkins\inst{\ref{ames}}
    \and D.W.~Latham\inst{\ref{cfa}}
    \and K.~Lester\inst{\ref{ames}}
    \and J.~McCormac\inst{\ref{warwick}}
    \and C.A.~Murray\inst{\ref{cavendish}}
    \and P.~Niraula\inst{\ref{miteaps}}
    \and P.P.~Pedersen\inst{\ref{cavendish}}
    \and D.~Queloz\inst{\ref{cavendish}}
    \and G.~Ricker\inst{\ref{mitkavli},\ref{mitphysics}}
    \and D.R.~Rodriguez\inst{\ref{stsci}}
    \and A.~Schroeder\inst{\ref{berkeley}}
    \and R.P.~Schwarz\inst{\ref{voorheesville}}
    \and N.~Scott\inst{\ref{ames}}
    \and S.~Seager\inst{\ref{miteaps},\ref{mitkavli},\ref{mitphysics},\ref{mitaeroastro}}
    \and C.A.~Theissen\inst{\ref{ucsd},\thanks{NASA Sagan Fellow}}
    \and S.~Thompson\inst{\ref{cavendish}}
    \and M.~Timmermans\inst{\ref{aru_liege}}
    \and J.D.~Twicken\inst{\ref{seti},\ref{ames}}
    \and J.N.~Winn\inst{\ref{princeton}}
}

\institute{
    Center for Space and Habitability, University of Bern, Gesellschaftsstrasse 6, 3012, Bern, Switzerland \label{unibe}
    \and Department of Earth, Atmospheric and Planetary Science, Massachusetts Institute of Technology, 77 Massachusetts Avenue, Cambridge, MA 02139, USA \label{miteaps}
    \and Kavli Institute for Astrophysics and Space Research, Massachusetts Institute of Technology, Cambridge, MA 02139, USA \label{mitkavli}
    \and Universidad Nacional de C\'ordoba - Observatorio Astron\'omico de C\'ordoba, Laprida 854, X5000BGR, C\'ordoba, Argentina \label{cordoba}
    \and Consejo Nacional de Investigaciones Cient\'ficas y T\'ecnicas (CONICET), Buenos Aires, Argentina \label{conicet}
    \and Universidad Nacional Aut\'onoma de M\'exico, Instituto de Astronom\'ia, AP 70-264, CDMX  04510, M\'exico \label{ciudad}
    \and Center for Astrophysics and Space Science, University of California San Diego, La Jolla, CA 92093, USA \label{ucsd}
    \and Max Planck Institute for Astronomy (MPIA), Koenigstuhl 17, D-69117 Heidelberg, Germany \label{mpia}
    \and Astrobiology Research Unit, Universit\'e de Li\`ege, All\'ee du 6 Ao\^ut 19C, B-4000 Li\`ege, Belgium \label{aru_liege}
    \and Space Sciences, Technologies and Astrophysics Research (STAR) Institute, Universit\' de Li\`ege, All\'ee du 6 Ao\^ut 19C, B-4000 Li\`ege, Belgium \label{star_liege}
    \and Department of Physics, Massachusetts Institute of Technology, Cambridge, MA 02139, USA \label{mitphysics}
    \and Universidad Nacional Aut\'onoma de M\'exico, Instituto de Astronom\'ia, AP 106, Ensenada 22800, BC, M\'exico \label{uname}
    \and Department of Physics \& Astronomy, Vanderbilt University, 6301 Stevenson Center Ln., Nashville, TN 37235, USA \label{vanderbilt}
    \and NASA Ames Research Center, Moffett Field, CA 94035, USA \label{ames}
    \and School of Physics \& Astronomy, University of Birmingham, Edgbaston, Birmimgham B15 2TT, UK \label{birmingham}
    \and Steward Observatory, The University of Arizona, 933 N. Cherry Avenue, Tucson, AZ 85721, USA \label{arizona}
    \and NASA Nexus For Exoplanetary System Science: Alien Earths Team \label{nexus}
    \and Lunar and Planetary Laboratory, The University of Arizona, 1639 E. University Boulevard, Tucson, AZ 85721, USA \label{lpl}
    \and Oukaimeden Observatory, High Energy Physics and Astrophysics Laboratory, Cadi Ayyad University, Marrakech, Morocco \label{oukaimeden}
    \and Graduate Institute of Astronomy, National Central University, 300 Jhongda Road, Zhongli, Taoyuan 32001, Taiwan \label{ncu_taiwan}
    \and Instituto de Astrof\'isica de Canarias (IAC), Calle V\'ia L\'actea s/n, 38200, La Laguna, Tenerife, Spain \label{iac}
    \and Center for Astrophysics | Harvard \& Smithsonian, 60 Garden Street, Cambridge, MA, 02138, USA \label{cfa}
    \and Dpto. F\'isica Te\'orica y del Cosmos, Universidad de Granada, 18071, Granada, Spain \label{granada}
    \and NASA Exoplanet Science Institute, Caltech/IPAC, Mail Code 100-22, 1200 E. California Blvd., Pasadena, CA 91125, USA \label{nesi}
    \and Vatican Observatory Research Group, University of Arizona, 933 N Cherry Ave., Tucson AZ, 85721-0065, USA \label{vat_arizona}
    \and Department of Astronomy, Tsinghua University, Beijing 100084, People's Republic of China \label{tsinghua}
    \and Department of Astronomy, 501 Campbell Hall, University of California at Berkeley, Berkeley, CA, 94720, USA \label{berkeley}
    \and Department of Physics \& Astronomy, University of California, Los Angeles, Los Angeles, CA 90095, USA \label{cali}
    \and Department of Physics, University of Warwick, Coventry, CV4 7AL, UK \label{warwick}
    \and Cavendish Laboratory, JJ Thomson Avenue, Cambridge, CB3 0HE, UK \label{cavendish}
    \and Space Telescope Science Institute, 3700 San Martin Drive, Baltimore, MD, 21218, USA \label{stsci}
    \and Patashnick Voorheesville Observatory, Voorheesville, NY 12186, USA \label{voorheesville}
    \and Department of Aeronautics and Astronautics, MIT, 77 Massachusetts Avenue, Cambridge, MA 02139, USA \label{mitaeroastro}
    \and SETI Institute, Mountain View, CA 94043, USA \label{seti}
    \and Department of Astrophysical Sciences, Princeton University, 4 Ivy Lane, Princeton, NJ 08544, USA \label{princeton}
}

\date{}

\abstract
    % context (optional)
    {Large sub-Neptunes are uncommon around the coolest stars in the Galaxy and are rarer still around those that are metal-poor. 
    However, owing to the large planet-to-star radius ratio, these planets are highly suitable for atmospheric study via transmission spectroscopy in the infrared, such as with JWST.}
    % aims (mandatory)
    {Here we report the discovery and validation of a sub-Neptune orbiting the thick-disk, mid-M dwarf star TOI-2406. 
    The star's low metallicity and the relatively large size and short period of the planet make TOI-2406\,b an unusual outcome of planet formation, and its characterisation provides an important observational constraint for formation models.}
    % methods (mandatory)
    {We first infer properties of the host star by analysing the star's near-infrared spectrum, spectral energy distribution, and Gaia parallax. 
    We use multi-band photometry to confirm that the transit event is on-target and achromatic, and we statistically validate the TESS signal as a transiting exoplanet.
    We then determine physical properties of the planet through global transit modelling of the TESS and ground-based time-series data.}
    % results (mandatory)
    {We determine the host to be a metal-poor M4~V star, located at a distance of 56\,pc, with properties $\mathrm{T_{eff} = 3100 \pm 75}$\,K, $\mathrm{M_* = 0.162 \pm 0.008}$\,$\mathrm{M_\odot}$, $\mathrm{R_* = 0.202 \pm 0.011}$\,$\mathrm{R_\odot}$, and $\mathrm{[Fe/H] = -0.38 \pm 0.07}$, and a member of the thick disk. 
    The planet is a relatively large sub-Neptune for the M-dwarf planet population, with $\mathrm{R_p = 2.94 \pm 0.17}$\,$\mathrm{R_\oplus}$ and $\mathrm{P = 3.077}$\,d, producing transits of 2\% depth. 
    We note the orbit has a non-zero eccentricity to 3$\mathrm{\sigma}$, prompting questions about the dynamical history of the system.}
    % conclusions (optional)
    {This system is an interesting outcome of planet formation and presents a benchmark for large-planet formation around metal-poor, low-mass stars. 
    The system warrants further study, in particular radial velocity follow-up to determine the planet mass and constrain possible bound companions. 
    Furthermore, TOI-2406\,b is a good target for future atmospheric study through transmission spectroscopy. 
    Although the planet's mass remains to be constrained, we estimate the S/N using a mass-radius relationship, ranking the system fifth in the population of large sub-Neptunes, with TOI-2406\,b having a much lower equilibrium temperature than other spectroscopically accessible members of this population. 
    }


\maketitle

%-------------------------------------------------------------------

\section{Introduction}
Short-period, rocky planets have been found to be common around M dwarfs; however, larger planets are rarer \citep{2015ApJ...807...45D,2015ApJ...798..112M,2019AJ....158...75H}. 
This extends down to the coolest stars, where only seven transiting large sub-Neptune ($2.75 < R_p < 4.0\,R_\oplus$) planets are known to orbit stars with effective temperatures below 3300\,K\footnote{\label{note:1}NASA Exoplanet Archive (04 March 2021)}.
The search for planets around these cool stars is ongoing with surveys such as SPECULOOS \citep{2018SPIE10700E..1ID,2020SPIE11445E..21S}, MEarth \citep{2008PASP..120..317N}, CARMENES \citep{2018A&A...612A..49R, Morales2019} and EDEN \citep{Gibbs2020}. 

From core accretion models \citep{1996Icar..124...62P}, the occurrence rate of large sub-Neptunes around late-M dwarfs is expected to be lower for metal-deficient environments \citep{Burn2021}. 
Therefore, if stellar metallicity is a good proxy for the amount of solid material in the protoplanetary disk, we would not expect to readily find large sub-Neptunes around metal-poor stars. 
In particular, planetesimal-based formation models do not readily produce these systems, and therefore their existence can be better explained by a pebble accretion scenario, in which the available solids can be transported to the planet from the outer disk. 

The large planet-to-star radius ratios for planets around cool stars means they produce deep transits. 
This makes them good targets for atmospheric study via transmission spectroscopy \citep{Seager2000, Brown2001}, particularly in the infrared, where M dwarfs are relatively bright. These wavelengths can be probed with HST and soon JWST. 
Notably, GJ~3470\,b (\citealt{2012A&A...546A..27B}; $\mathrm{T_{eff}}$ = 3600\,K, $\mathrm{R_p}$ = 4.6\,$\mathrm{R_\oplus}$) and GJ~1214\,b (\citealt{Charbonneau2009}; $\mathrm{T_{eff}}$ = 3026\,K, $\mathrm{R_p}$ = 2.8\,$\mathrm{R_\oplus}$) are two of the most well-studied planets by transmission spectroscopy\textsuperscript{\ref{note:1}}. 

Here we report the discovery and validation of a sub-Neptune planet orbiting a metal-poor M4~V host star ($\mathrm{T_{eff}}$ = 3100\,K, $\mathrm{R_p}$ = 2.9\,$\mathrm{R_\oplus}$). 
In Sect.~\ref{sec:obs} we detail the TESS and follow-up observations. 
In Sect.~\ref{sec:stellar} we present the stellar characterisation analysis. 
Then in Sect.~\ref{sec:validation}, owing to improved precision over TESS from the ground, we validate the transit signal as a bona fide planet.
We present our search for further planets in the TESS data in Sect.~\ref{sec:search}, before giving the results from our transit modelling of the photometric light curves in Sect.~\ref{sec:mcmc}. 
Finally, in Sect.~\ref{sec:discussion} we discuss the system's formation, architecture, and potential for follow-up study, and we conclude in Sect.~\ref{sec:conclusion}.

%-----------------------------------------------------------------

\section{Observations} \label{sec:obs}

\subsection{TESS photometry}
\begin{figure*}
\centering
\subfigure{\includegraphics[width=0.45\textwidth]{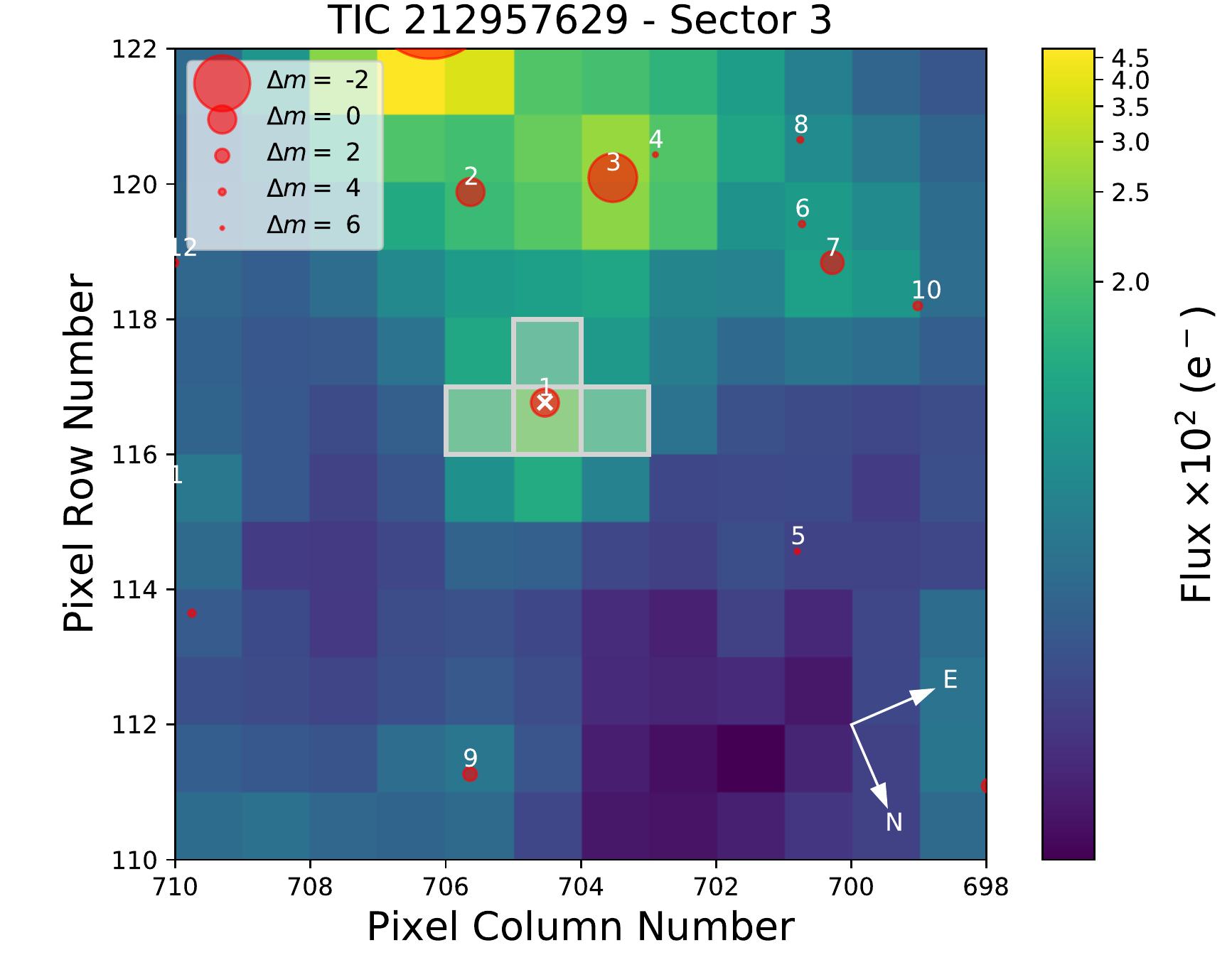}} 
\quad
\subfigure{\includegraphics[width=0.45\textwidth]{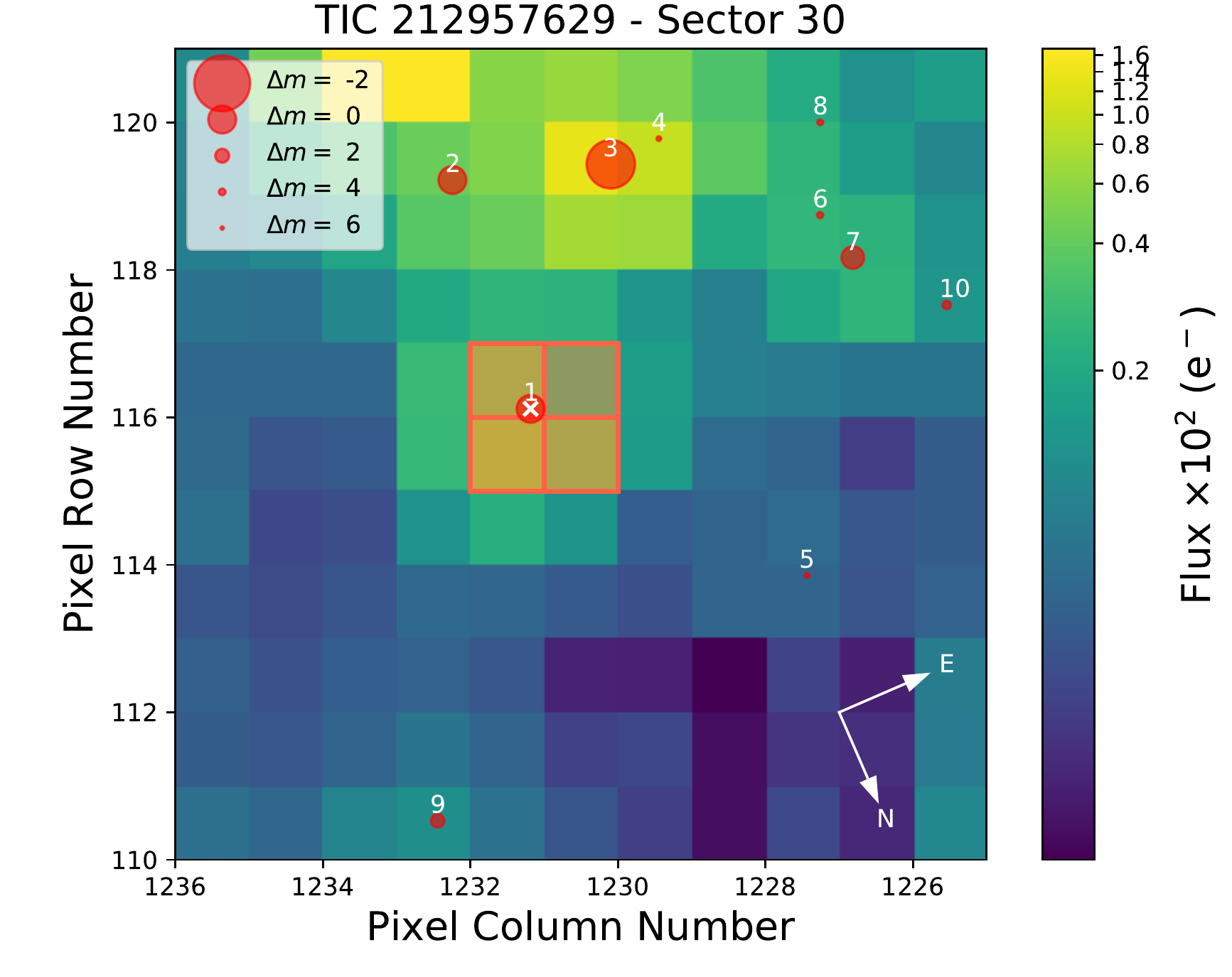}} 
\caption{Target pixel file (TPF) images for TOI-2406, from TESS Sectors 3 (left) and 30 (right). The apertures used for light curve generation are over-plotted; sources identified in Gaia DR2 are also included, with symbols correlated to their brightness compared to the target. These images were produced using \texttt{tpfplotter} \citep{2020A&A...635A.128A}.}
\label{fig:tpf}
\end{figure*}

TOI-2406 (TIC~212957629) was first observed by TESS \citep{2015JATIS...1a4003R} in the full frame images (FFIs) at 30-minute cadence in Sector 3, between 20 September and 18 October 2018. 
It was observed by TESS again two years later, at 2-minute cadence in Sector 30, between 22 September and 21 October 2020. 
The two sectors are shown in Fig.~\ref{fig:tpf}, along with the nearby Gaia sources \citep{GaiaDR2}. 
The star is present in the TESS Input Catalogue v8.1 (TIC; \citealt{2019AJ....158..138S}), identified as an M4.5 dwarf located at 56\,pc. 
The Science Processing Operations Center (SPOC; \citealt{2016SPIE.9913E..3EJ}) at NASA Ames Research Center extracted the photometry for this target from the 2-minute data from Sector 30, and conducted a transiting planet search \citep{2002ApJ...575..493J,2010SPIE.7740E..0DJ} on 30 October 2020. This yielded a strong 2\% deep transit-like signature at a period of 3.077\,d. 
The TESS Object of Interest (TOI) vetting team at MIT reviewed the SPOC Data Validation reports \citep{2018PASP..130f4502T,2019PASP..131b4506L} for this signal and released TOI-2406 on 24 November 2020 \citep{Guerrero2021}. 

We retrieved the TESS Sector 3 FFI light curve via point spread function (PSF) photometry on a $11 {\times} 11$ pixel region around the target, which we downloaded via TESSCut \citep{2019ascl.soft05007B}. In order to perform photometry, we retrieved the source position from TICv8 and corrected it for proper motion. We then interpolated the empirical, super-sampled pixel response function\footnote{\url{https://archive.stsci.edu/tess/all_products.html}} (PRF) using a third-order polynomial. A flat template for the background sky emission was used. Finally, we determined the fluxes (i.e. template coefficients) of the target and the sky background for all time bins via linear regression.
The data release notes for Sector 3\footnote{\url{https://archive.stsci.edu/missions/tess/doc/tess_drn/tess_sector_03_drn04_v02.pdf}} indicate that TESS conducted a series of engineering tests during its usual data gap (the perigee orbit) to improve the pointing accuracy. We inspected the data to confirm that its quality is reliable and included it in our analysis. 

\begin{figure*}
\centering
\includegraphics[width=17cm]{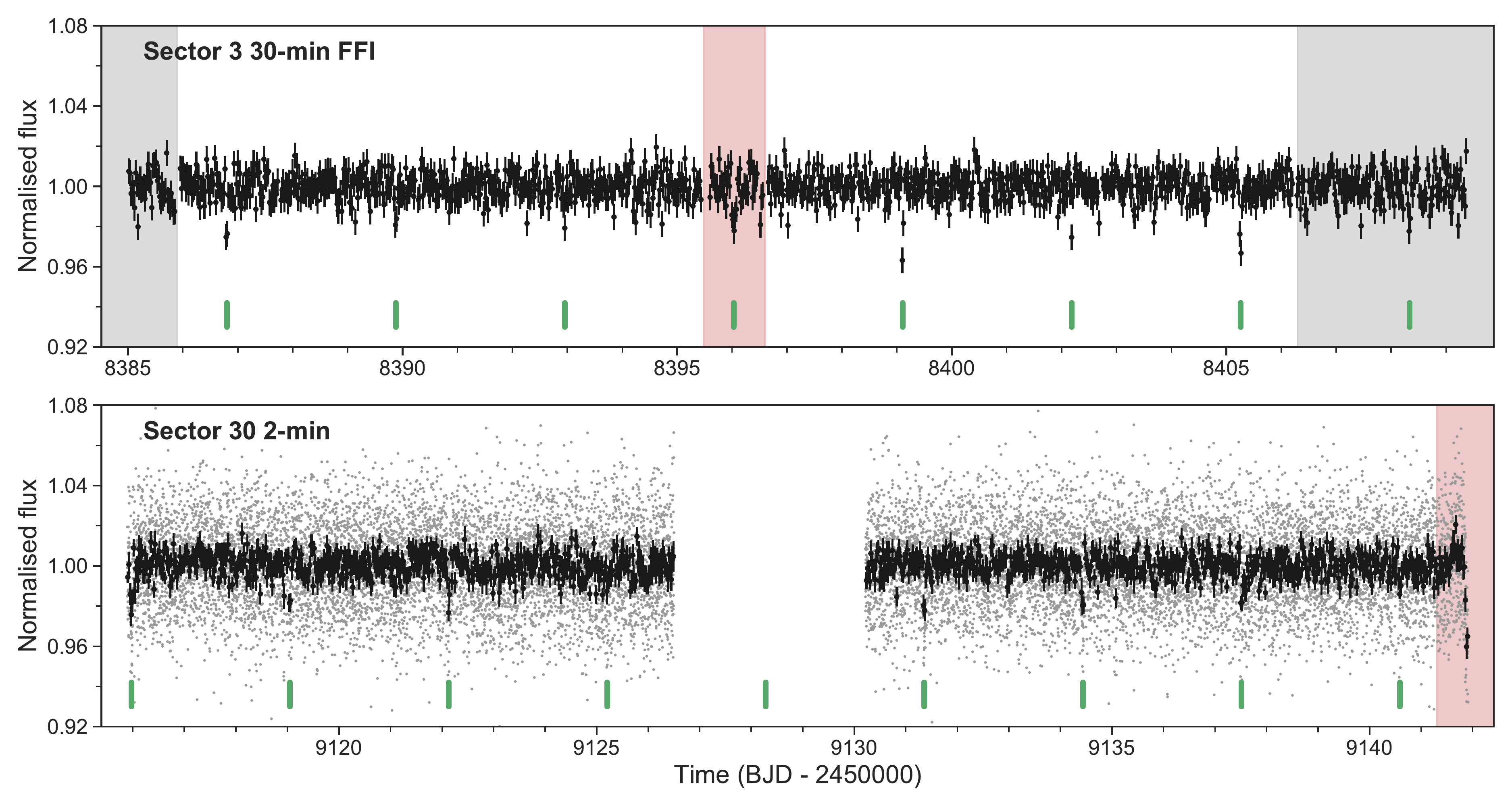}
\caption{Sector 3 light curve created from FFIs (top) and the Sector 30 2-minute light curve (bottom). 
The 2-minute data points (grey) have been binned by 15 to produce the black points, with error bars corresponding to the standard deviation in the bins, which match the FFI light curve cadence of 30 min. 
Regions marked in red were excluded from the analysis: In Sector 3, there were poor-quality engineering data between orbits; and in Sector 30, a significant trend at the end was possibly caused by scattered light. 
Regions marked in grey correspond to data taken outside of science data acquisition for testing purposes but used in the analysis. 
One transit occurred in this period, towards the end of Sector 3. 
The transits of TOI-2406\,b are labelled with green markers below the light curves.}
\label{fig:tess-data}
\end{figure*}

We downloaded the TESS Sector 30 SPOC Presearch Data Conditioning Simple Aperture Photometry (PDC-SAP; \citealt{2012PASP..124..985S,2014PASP..126..100S,2012PASP..124.1000S}) light curve from MAST. 
Long-term trends are already removed in the PDC light curve. We assessed further detrending via a median filter but found no improvement to the light curve quality, likely because it is dominated by white noise.
All data with a non-zero quality flags were removed, along with the final $\sim$1\,d of the sector. 
We also tested removing the first transit of the sector, but found no change in our Markov chain Monte Carlo (MCMC) results, and therefore included these data in our analysis.
Both TESS light curves are displayed in Fig.~\ref{fig:tess-data}, showing the transit times and regions excluded from the transit analysis. 

\subsection{Follow-up photometry}
We obtained ground-based follow-up of multiple transit events, coordinated by the TESS Follow-up Observing Program (TFOP) sub-group~1 for Seeing-limited Photometry. 
The observations are summarised in Table~\ref{table:observations} and detailed in the following sections. 

\begin{table*}
\caption{Summary of photometric observations. Shown are: the date(s) of the observations; the number of transits covered; the telescope, filter, and exposure time used; and the RMS of the 30-minute binned light curve.} 
\label{table:observations}
\centering
\begin{tabular}{l c c c c c c}
\hline\hline
Night of (UT) & $\mathrm{N_{tr}}$ & Telescope & Filter & Exposure (s) & $\mathrm{\sigma_{30min}}$ (ppt) \\[0.1cm]
\hline
20 Sep -- 18 Oct 2018 & 6 & TESS Sector 3 & TESS & 1800 & 6.5 \\[0.1cm]
22 Sep -- 21 Oct 2020 & 8 & TESS Sector 30 & TESS & 120 & 5.7 \\[0.1cm]
29 Nov 2020 & 1 & TRAPPIST-South & Exo & 120 & 1.4 \\[0.1cm]
02 Dec 2020 & 1 & SAINT-EX & $\mathrm{z^\prime}$ & 120 & 0.5 \\[0.1cm]
05 Jan 2021 & <1 & TRAPPIST-South & Exo & 120 & -- \\[0.1cm]
05 Jan 2021 & 1 & EDEN/VATT & V & 120 & 2.6 \\[0.1cm]
09 Jan 2021 & 1 & LCO-McDonald & $\mathrm{i^\prime}$ & 150 & 0.9 \\[0.1cm]
\hline
\end{tabular}
\end{table*}

\subsubsection{TRAPPIST-South}
TOI-2406 was first followed-up on 29 November 2020 with the 0.6 m TRAPPIST-South telescope, located at La Silla Observatory, Chile \citep{2011Msngr.145....2J,2011EPJWC..1106002G}. 
A full transit was observed in the Exo bandpass (500-1100\,nm), consisting of 74 images each with an exposure time of 120\,s, covering 168 minutes. 
The data were reduced, and a light curve was produced using the AstroImageJ (AIJ) software \citep{2017AJ....153...77C}. 
A partial transit was also observed on 05 January 2021; however, this was not used in our analysis due to the small amount of transit coverage. 

\subsubsection{SAINT-EX}
We obtained a full transit with SAINT-EX on 02 December 2020 in $\mathrm{z^\prime}$. 
SAINT-EX (Search And characterIsatioN of Transiting EXoplanets) is a 1-m telescope located at the Observatorio Astron\'omico Nacional, San Pedro M\'artir, Mexico \citep{2018csss.confE..59S}. 
The observations consisted of 79 images with an exposure time of 120\,s, covering 178 minutes total. 
The data were reduced using a custom pipeline, PRINCE, which is detailed in \citet{2020A&A...642A..49D}. 
The final light curve was produced by a weighted principal component analysis (PCA) approach \citep{bailey2012}\footnote{\url{https://github.com/jakevdp/wpca}} to correct for common trends between the target and comparison stars. 
We also produced a light curve using AIJ, but found it to have a higher out-of-transit scatter and a larger trend. We therefore used the PRINCE light curve in our analysis.
The SAINT-EX light curve had a 30-minute rms of $\sim$500\,ppm (see Table~\ref{table:observations}), with a good amount of baseline on either side of transit, making this our most constraining dataset. 

\subsubsection{VATT}
We also obtained a $V$-band transit with the VATT4K CCD Imager on the 1.8 m Vatican Advanced Technology Telescope (VATT) on 06 January 2021 (UT).
The data were collected as part of Project EDEN \citep[Exo-Earth Discovery and Exploration Network; e.g.][]{Gibbs2020}.
We collected 88 images each with an exposure time of 120\,s over 201 minutes, covering almost the full transit. 
We initially followed the data reduction and calibration procedure for Project EDEN observations, outlined by \citet{Gibbs2020}. 
We then also produced a light curve using AIJ for aperture photometry and weighted PCA for differential photometry, which we found to be improved over the initial pipeline reduction, and therefore we used this in our analysis. 

\subsubsection{Las Cumbres Observatory (LCO)}
A final transit was observed on 09 January 2021 with the Las Cumbres Observatory Global Telescope \citep[LCOGT;][]{2013PASP..125.1031B}. 
The observations were taken using the 1-m telescope at McDonald Observatory, using the $\mathrm{i^\prime}$ filter. 
We took 55 exposures of 150\,s over 173 minutes, covering the full transit, although lacking some post-transit baseline. 
The data were reduced, and a light curve was produced using AIJ.

\subsection{Spectroscopy}
Infrared spectroscopy of TOI-2406 was obtained with the SpeX spectrograph \citep{2003PASP..115..362R} on the 3.2-m NASA Infrared Telescope Facility on Maunakea, Hawaii, on 03 December 2020 (UT). Conditions were clear with 0$\farcs$4 seeing. The short-wavelength cross-dispersed (SXD) mode was used with the 0$\farcs$3-wide slit to obtain a 0.7--2.5~$\mu$m spectrum in seven orders at a spectral resolving power $\lambda/\Delta\lambda$ $\approx$ 2000. A total of four ABBA nod sequences (16 exposures) were obtained with an integration time of 170~s per exposure with the slit aligned with the parallactic angle. The A0~V star HD 13936 ($V$ = 6.549) was observed afterwards at an equivalent airmass for flux and telluric calibration, followed by arc lamp and flat field lamp exposures. Data were reduced using SpeXtool v4.1 \citep{2004PASP..116..362C} using standard settings. The resulting spectrum of TOI-2406 had a median signal-to noise ratio (S/N) of 60, with JHK peaks of around 100-150.

\begin{figure}[tb!]
\centering
\includegraphics[width=0.48\textwidth]{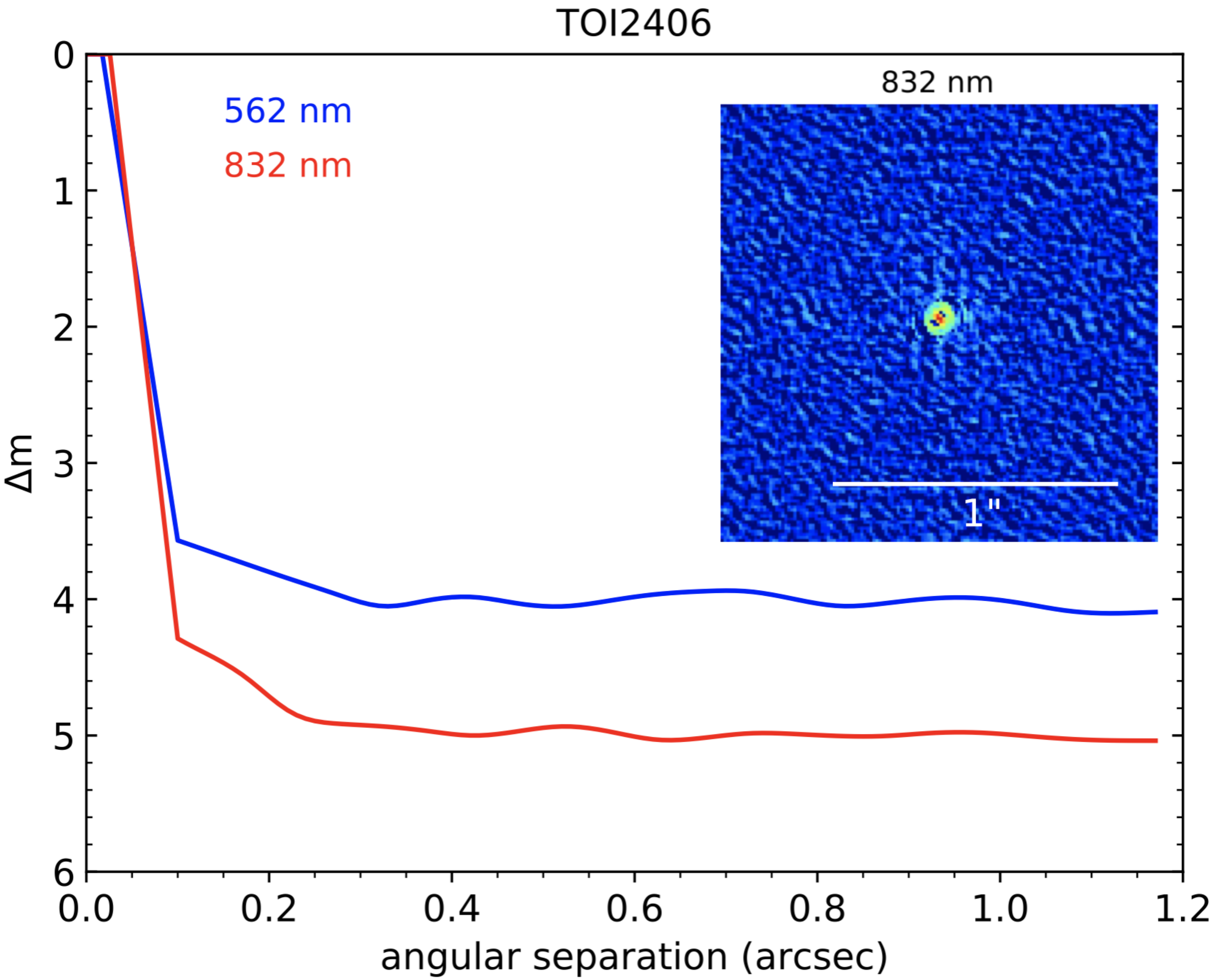}
\caption{Zorro speckle imaging 5$\sigma$ contrast curves, along with the reconstructed 832 nm image.}
\label{fig:speckle}
\end{figure}

\subsection{High-resolution imaging}
TOI-2406 was observed with the Zorro instrument mounted on the 8-metre Gemini-South telescope on 29 December 2020. 
Zorro simultaneously observes in blue (562/54\,nm) and red (832/40\,nm) bandpasses, with inner working angles of 0$\farcs$026 in the blue and 0$\farcs$017 in the red. 
Three thousand 0.06-s images were obtained and combined in the data reduction process
and the Fourier analysis as described in \citet{2011AJ....142...19H}. 
The Zorro observations do not reveal a previously unknown companion to TOI-2406 within the 5$\sigma$ contrast limits obtained (see Fig.~\ref{fig:speckle}). 
The contrast curve begins to flatten at $\sim$0$\farcs$1 and appears flat from $\sim$0$\farcs$2. Given the distance to the target of 55.6\,pc, these correspond to projected distances of 5.6 and 11.1\,AU away from the star, respectively. 
The observations had a contrast of 4--5\,magnitudes beyond 0$\farcs$2, allowing us to rule out companion stars earlier than about spectral type M8 beyond this separation. 
We also find no evidence for companion stars within 0$\farcs$2 in the near-infrared spectrum. 
Therefore, any possible companions would be unable to explain the 2\% transit depth seen in the TESS and follow-up data.

%-----------------------------------------------------------------

\section{Stellar properties} \label{sec:stellar}
TOI-2406 is a quiet, mid-M dwarf located towards the ecliptic plane ($l = -6.6^\circ$) at 56\,pc. As we show below from an analysis of the space velocity, it is very likely a member of the Galactic thick-disk population, and therefore would have an age of $11 \pm 1.5$ Gyr \citep{2021A&A...645A..85M}. 
The star is faint in the optical (V$\gtrsim$17) but reasonably bright in the near-infrared (J$=$12.6). 
The catalogued astrometric and photometric parameters of TOI-2406 are given in Table~\ref{table:mags}. 
In this section we detail the methodology used to determine the properties of the star, given in Table~\ref{table:stellar}. 

\begin{table}
\caption{Astrometric and photometric properties of the host star.
1: Gaia EDR3 \citep{2020arXiv201201533G}; 2: this work; 3: TIC \citep{2019AJ....158..138S}; 4: Pan-STARRS \citep{2016arXiv161205560C}; 5: 2MASS \citep{2006AJ....131.1163S}; 6: WISE \citep{2014yCat.2328....0C}.} 
\label{table:mags}
\centering
\begin{tabular}{l c c}
\hline\hline
Property & Value & Source \\[0.1cm]
\hline
Identifiers: &  &  \\[0.1cm]
TOI & 2406 &  \\[0.1cm]
TIC & 212957629 &  \\[0.1cm]
LP & 645-50 &  \\[0.1cm]
2MASS & J00351318-0322140 &  \\[0.1cm]
Gaia ID & 2528453161326406016 &  \\[0.4cm]

Astrometry: &  &  \\[0.1cm]
RA (J2000) & 00:35:13.22 & [1] \\[0.1cm]
Dec (J2000) & $-$03:22:14.29 & [1] \\[0.1cm]
$\mathrm{\mu_{RA}}$ (mas/yr) & $226.01 \pm 0.05$ & [1] \\[0.1cm]
$\mathrm{\mu_{Dec}}$ (mas/yr) & $-336.24 \pm 0.04$ & [1] \\[0.1cm]
Parallax (mas) & $17.98 \pm 0.041$ & [1] \\[0.1cm]
Distance (pc) & $55.60 \pm 0.13$ & [1] \\[0.1cm]
V$_{tan}$ (km/s) & 106.8 $\pm$ 0.2 & [1] \\[0.1cm]
RV (km/s) & +15 $\pm$ 6 & [2] \\[0.1cm]
U (km/s) & +1.5 $\pm$ 0.9 & [2] \\[0.1cm]
V (km/s) & $-$93.2 $\pm$ 2.1 & [2] \\[0.1cm]
W (km/s) & $-$46.4 $\pm$ 5.2 & [2] \\[0.4cm]

Photometry: &  &  \\[0.1cm]
TESS (mag) & 14.31 & [3] \\[0.1cm]
% Gaia EDR3 - Vizier
BP (mag) & $17.434 \pm 0.007$ & [1] \\[0.1cm]
G (mag) & $15.663 \pm 0.003$ & [1] \\[0.1cm]
RP (mag) & $14.393 \pm 0.004$ & [1] \\[0.1cm]
% Pan-STARRS1
g (mag) & $17.7198 \pm 0.0081$ & [4] \\[0.1cm]
r (mag) & $16.5656 \pm 0.0037$ & [4] \\[0.1cm]
i (mag) & $14.9479 \pm 0.0084$ & [4] \\[0.1cm]
z (mag) & $14.2083 \pm 0.0037$ & [4] \\[0.1cm]
y (mag) & $13.8748 \pm 0.0049$ & [4] \\[0.1cm]
% 2MASS
J (mag) & $12.633 \pm 0.024$ & [5] \\[0.1cm]
H (mag) & $12.129 \pm 0.024$ & [5] \\[0.1cm]
K (mag) & $11.894 \pm 0.025$ & [5] \\[0.1cm]
% WISE
W1 (mag) & $11.706 \pm 0.023$ & [6] \\[0.1cm]
W2 (mag) & $11.458 \pm 0.022$ & [6] \\[0.1cm]
W3 (mag) & $11.049 \pm 0.143$ & [6] \\[0.1cm]
W4 (mag) & $> 8.593$ & [6] \\[0.1cm]
\hline
\end{tabular}
\end{table}

\subsection{Spectroscopy} \label{ref_spectro}

\begin{figure}
\centering
\includegraphics[width=0.48\textwidth]{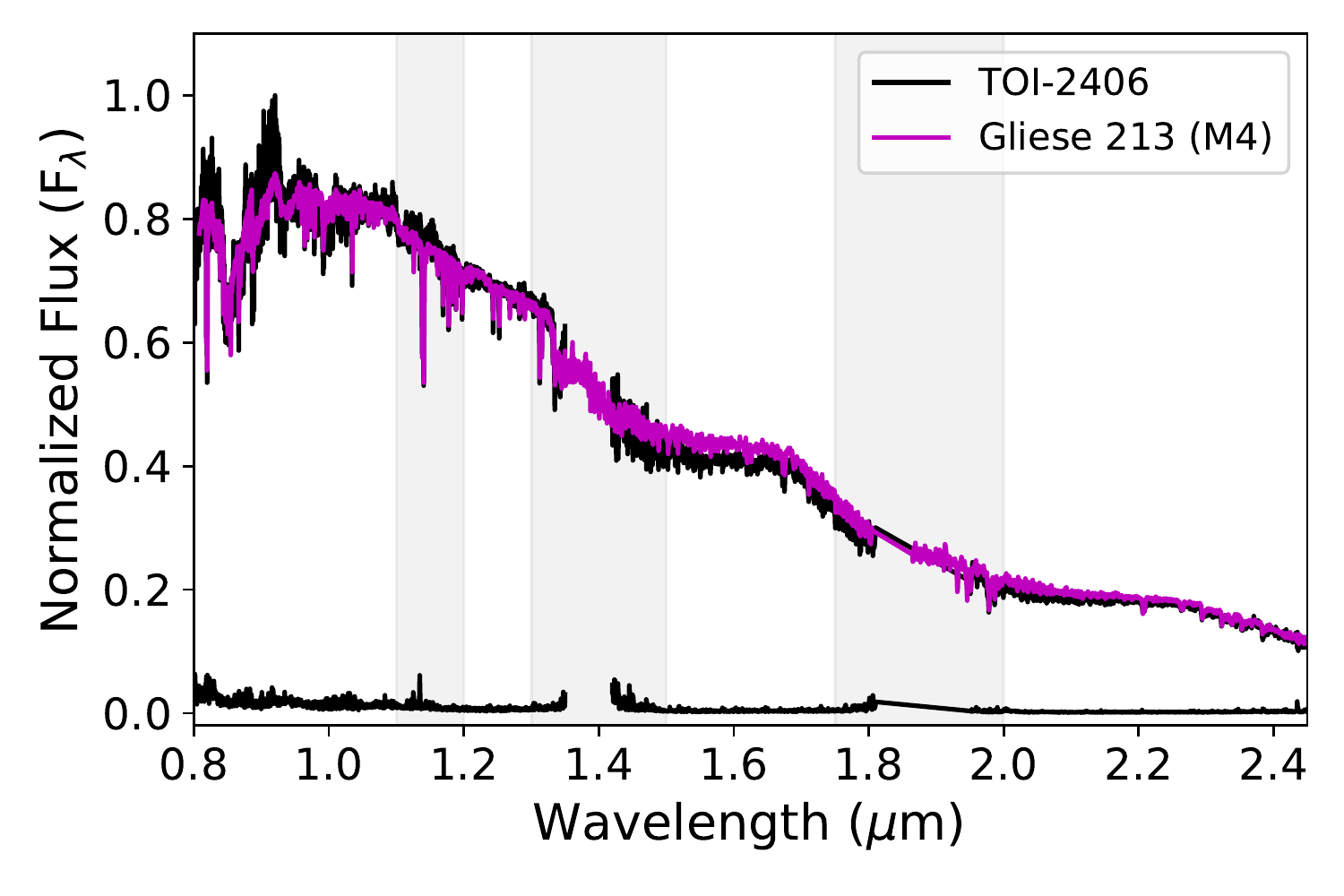} 
 
\includegraphics[width=0.48\textwidth]{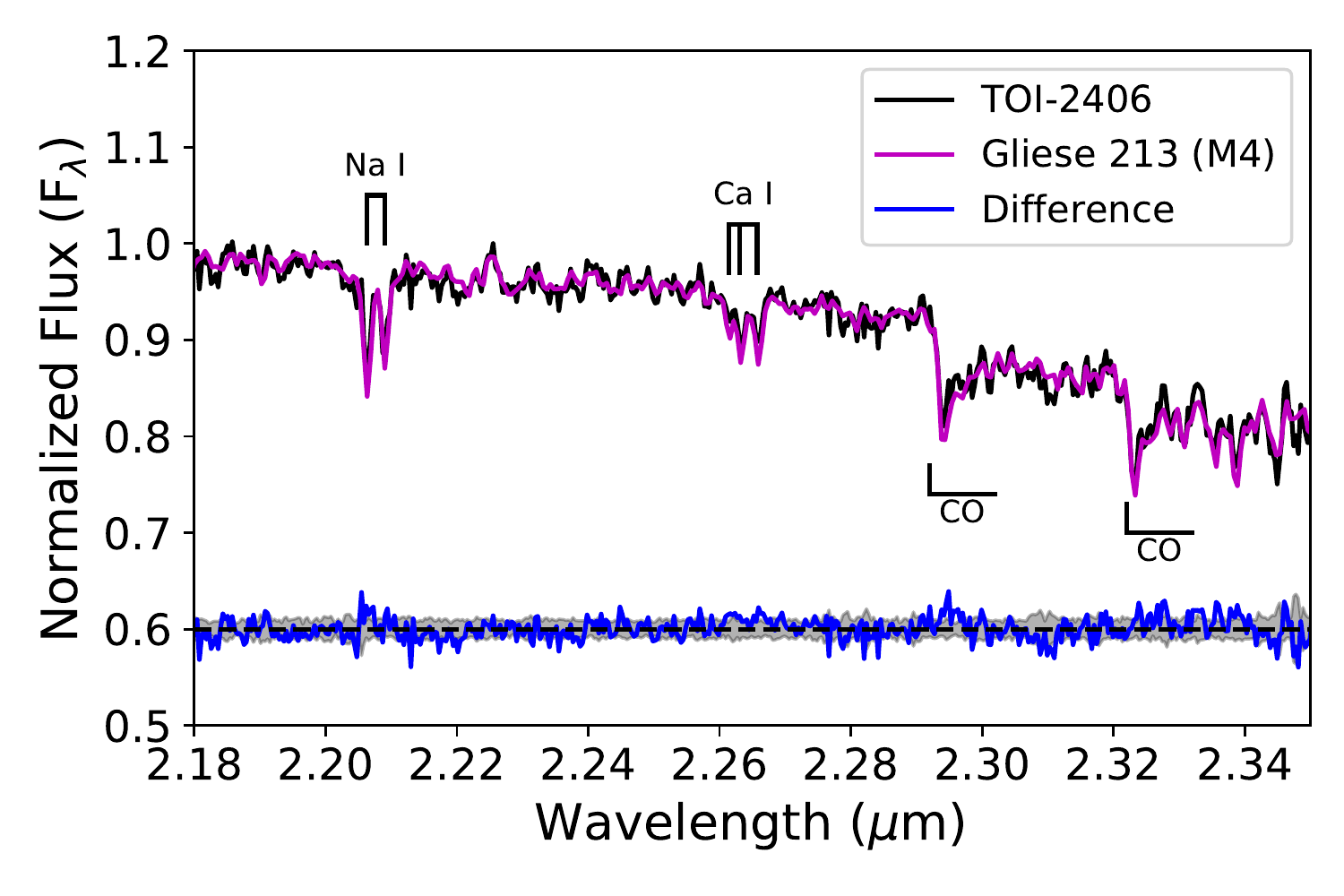}  
\caption{SpeX/SXD near-infrared spectrum of TOI-2406 (black line) compared to equivalent data for the M4.0 spectral standard Gliese 213 (data from \citealt{2009ApJS..185..289R}; magenta line). Top: Full spectra, with the uncertainty spectrum of TOI-2406 is shown in black along the bottom, and regions of strong telluric absorption are indicated by the grey panels.
Bottom: Close-up of the K-band region for these two spectra, highlighting the metallicity-sensitive absorption features Na~I ($\lambda$2.2079$\mu$m) and Ca~I ($\lambda$2.2640$\mu$m), as well as CO band heads. The difference between the spectra (blue line) is consistent with the measurement uncertainties (grey band).}
\label{fig:spex}
\end{figure}

The near-infrared spectrum of TOI-2406 is shown in Figure~\ref{fig:spex} in comparison to equivalent data for the M4 spectral standard Gliese 213 from the IRTF Spectral Library \citep{2009ApJS..185..289R}. Both the overall infrared spectral energy distributions (SEDs) and the detailed line features in these spectra are well matched. 
We evaluated the metallicity-sensitive lines Na~I (2.2079\,$\mu$m) and Ca~I (2.2640\,$\mu$m) using the empirical calibrations of \citet{2013AJ....145...52M} to infer an average metallicity of [Fe/H] = $-$0.38$\pm$0.07 dex (i.e. significantly metal-poor compared to the Sun). We also used the line centres for Na~I, Mg~I, Al~I, K~I, and Ca~I to infer a heliocentric velocity of +15$\pm$6~km/s. Combining this with the tangential velocity from {\it Gaia\/} Early Data Release 3 \citep[EDR3;][]{2020arXiv201201533G}, we infer the local standard of rest (LSR) $UVW$ velocities listed in Table~\ref{table:mags}. 

The significant negative $V$ velocity relative to the LSR makes this source a likely thick-disk star, based on the kinematic sample of \citet{2003AA...410..527B}. 
Following the methodology of \citet{2014A&A...562A..71B}, we compute membership probabilities of 0.3, 99.5, and 0.2 per cent for the thin-disk, thick-disk, and halo populations, respectively. 
Therefore, we reason that TOI-2406 is a member of the thick disk, which is consistent with the inferred sub-solar metallicity.  
We note that Gliese 213 is also a high-velocity star with thick-disk kinematics and a slightly sub-solar metallicity based on its near-infrared spectrum ([Fe/H] = -0.27$\pm$0.05 dex).

\subsection{SED fitting, empirical relations, and evolutionary modelling} \label{sec:sedstellar}
\begin{figure}
    \centering
    \includegraphics[width=\linewidth,trim=80 60 90 70,clip]{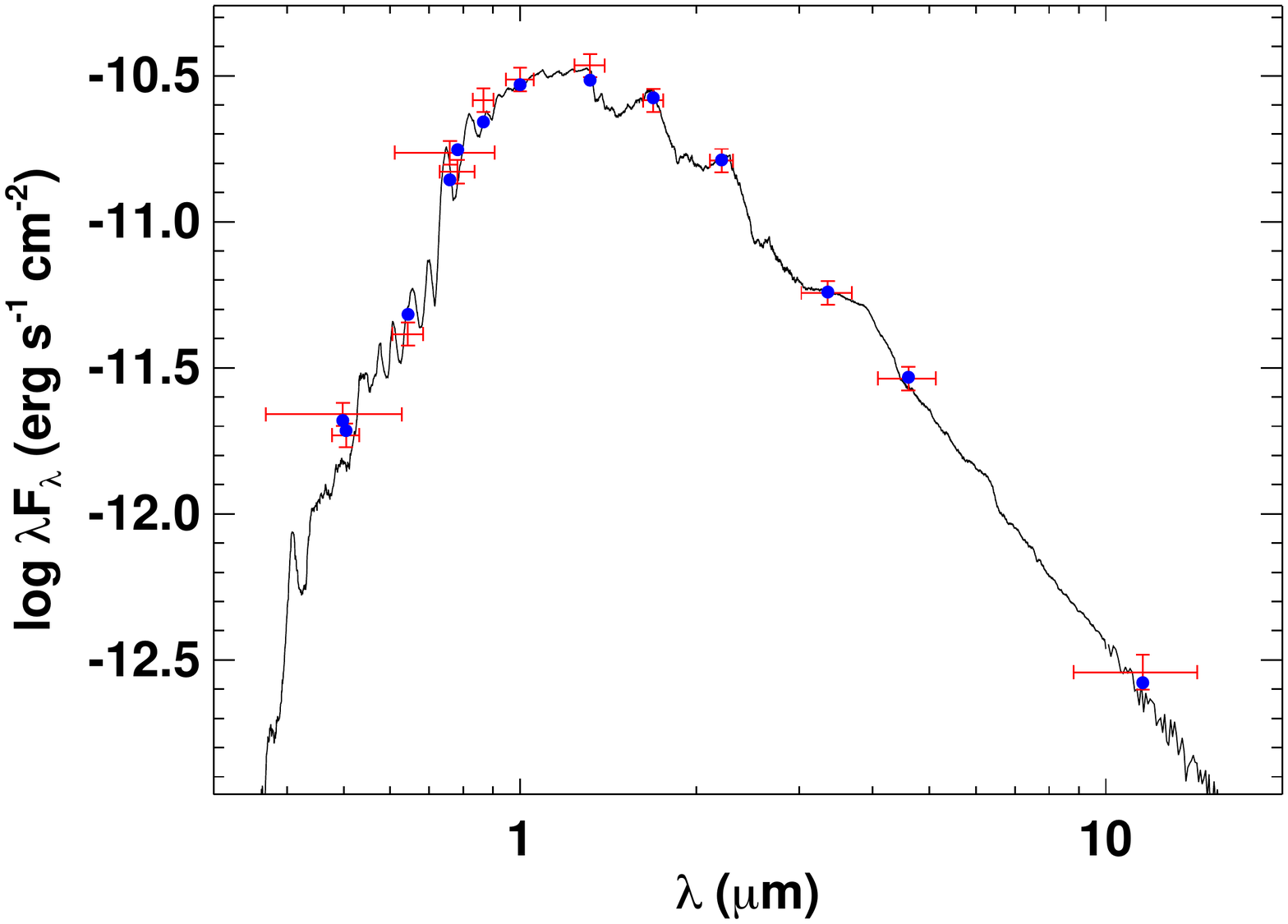}
    \caption{SED fit of TOI-2406. The black curve is the best fitting BT-Dusty model, red symbols are the observed fluxes (horizontal bars represent the effective bandpass widths), and blue symbols are the model fluxes.}
    \label{fig:sed}
\end{figure}

We performed an analysis of the broadband SED of the star together with the {\it Gaia\/} EDR3 parallax \citep[with no systematic offset applied; see e.g.][]{StassunTorres:2021}, in order to determine an empirical measurement of the stellar radius, following the procedures described in \citet{Stassun:2016,Stassun:2017,Stassun:2018}. 
We pulled the $grizy$ magnitudes from {\it Pan-STARRS}, the $JHK_S$ magnitudes from {\it 2MASS}, and the W1--W3 magnitudes from {\it WISE}. 
We opted not to use the $G, \,G_{\rm BP}, \,G_{\rm RP}$ magnitudes from {\it Gaia}, as these very broad filters are less ideal than the narrower bandpasses that are available. 
Together, the available photometry spans the stellar SED over the wavelength range 0.4--10~$\mu$m (see Fig.~\ref{fig:sed}).

We performed a fit using the BT-Dusty stellar atmosphere models \citep{2012RSPTA.370.2765A}, fitting for the effective temperature ($T_{\rm eff}$), metallicity ([Fe/H]), and surface gravity ($\log g$). 
The extinction ($A_V$) was fixed at zero due to the star's proximity. 
The resulting fit, with $T_{\rm eff} = 3100 \pm 75$~K, $\log g = 5.0 \pm 0.5$, and [Fe/H] $= -0.5 \pm 0.5$, is reasonably good (Fig.~\ref{fig:sed}), with a reduced $\chi^2$ of 1.4. 
Integrating the (un-reddened) model SED gives the bolometric flux at Earth, $F_{\rm bol} = 3.53 \pm 0.17 \times 10^{-11}$ erg~s$^{-1}$~cm$^{-2}$. 
Taking the $F_{\rm bol}$ and $T_{\rm eff}$, together with the {\it Gaia\/} parallax, gives the stellar radius, $R_\star = 0.202 \pm 0.011$~R$_\odot$. 

To assess the reliability of our stellar radius measurement, we compared our SED model result to empirical relations for cool stars. 
Both the \citet{2015ApJ...804...64M} $\mathrm{M_K{-}R_*}$ relation and the \citet{2012ApJ...757..112B} $\mathrm{M_*{-}R_*}$ relation (see discussion of $\mathrm{M_*}$ in the following paragraphs) return a stellar radius of $0.197 \pm 0.010\,R_\odot$. 
This is in good agreement with our SED radius (0.5$\sigma$), which we use as our final stellar radius estimate.

For the mass, we applied stellar evolutionary modelling, using the models for very low-mass stars presented in \citet{2019ApJ...879...94F}. 
We used as constraints the luminosity derived from $F_{\rm bol}$ and the {\it Gaia\/} EDR3 parallax ($L_\star = 3.44 \pm 0.16 \times 10^{-3} L_{\odot}$), the metallicity inferred in Sect. \ref{ref_spectro}, and assuming an age of $\gtrsim$ 2 Gyr, in the absence of signs of a young star (e.g. no fast rotation and no flares seen in the light curves; Sect.~\ref{sec:obs})\footnote{The luminosity of very low-mass stars evolves very slowly with time once the star has turned on core H-burning and has reached the main sequence.}. 
We obtained a stellar mass of 0.163$\pm$0.007 $M_{\odot}$. 
This uncertainty reflects the error propagation on the stellar luminosity and metallicity, but also the uncertainty associated with the input physics of the stellar models \citep{2018ApJ...853...30V}. 

We also estimated the mass through the empirical relations, the mass-$M_K$ relation of \citet{Mann:2019} and the mass-luminosity relation of \citet{2016AJ....152..141B}, finding values of $0.165 \pm 0.008$ and $0.158 \pm 0.008$ $\mathrm{M_*}$, respectively.
Combining these empirical estimates and our evolutionary modelling value by a simple average, we obtain our final stellar mass estimate of 0.162$\pm$0.008 $M_{\odot}$.
Considering the stellar radius estimate from SED fitting, this gives a mean stellar density of $\rho_\star = 27.7_{-4.2}^{+5.3}$~g~cm$^{-3}$.

\begin{table}
\caption{Properties of the host star. Metallicity is based on the empirical calibrations of \citet{2014AJ....147..160M} and \citet{2014AJ....147...20N}.} 
\label{table:stellar}
\centering
\begin{tabular}{l c c}
\hline\hline
Property & Value & Source \\[0.1cm]
\hline
Sp.\ type & M4 & Spectrum \\[0.1cm]
$\mathrm{T_{eff}}$ (K) & $3100 \pm 75$ & SED \\[0.1cm]
$\mathrm{[Fe/H]}$ & $-0.38 \pm 0.07$ & Spectrum \\[0.1cm]
$\mathrm{M_*}$ ($\mathrm{M_\odot}$) & $0.162 \pm 0.008$ & See text \\[0.1cm]
$\mathrm{R_*}$ ($\mathrm{R_\odot}$) & $0.202 \pm 0.011$ & SED \\[0.1cm]
$\mathrm{F_{bol}}$ ($10^{-11}$ $\mathrm{erg\,s^{-1}\,cm^{-2}}$) & $3.53 \pm 0.17$ & SED \\[0.1cm]
$\mathrm{L_*}$ ($10^{-3}$ $\mathrm{L_\odot}$) & $3.40 \pm 0.16$ & SED \\[0.1cm]
$\mathrm{\log\,g}$ & $5.037_{-0.051}^{+0.053}$ & $\mathrm{M_*}$, $\mathrm{R_*}$ \\[0.1cm]
$\mathrm{\rho_*}$ ($\mathrm{g\,cm^{-3}}$) & $27.7_{-4.2}^{+5.3}$ & $\mathrm{M_*}$, $\mathrm{R_*}$ \\[0.1cm]
\hline
\end{tabular}
\end{table}

%-----------------------------------------------------------------

\section{Planet validation} \label{sec:validation}

\subsection{TESS Data Validation Report}

We initially vetted the target before obtaining further observations by examining the TESS Data Validation Report for Sector 30 \citep{2018PASP..130f4502T,2019PASP..131b4506L}. 
TOI-2406.01 was summarised as a strong periodic transit signal with a phase-folded S/N of 13.2. 
The odd and even transit depths agreed to within $0.6 \sigma$, and there was no apparent secondary eclipse. 
TESS uses a large spatial pixel scale of 20\arcsec, making the aperture of a single star $\sim$1\arcmin, which could be contaminated by light from nearby stars. 
The star has no resolved neighbours within 1\,arcmin, and those nearest are faint and do not provide enough light in the aperture to produce the observed 2\% deep signal.
All other tests, including centroid offset and ghost diagnostic, were also passed.

\subsection{Follow-up photometry}
The ground-based follow-up photometry allowed a two-fold test of false-positive scenarios: first, by confirming the TESS signal was on the expected star, and second, by assessing the transit depth at multiple wavelengths. The follow-up photometry also has much-improved precision over the TESS data, in particular the SAINT-EX light curve, allowing a more stringent test of the transit shape (see Sect.~\ref{sec:fpp}). 

In contrast to the TESS large pixel scale, the follow-up photometry we obtained were extracted with apertures of a few arcseconds. 
No dimming events were seen on any nearby stars, while we detected clear 2\% transits on the target in each follow-up light curve. 

Our photometric observations comprised five different bandpasses -- TESS, V, $\mathrm{i^\prime}$, $\mathrm{z^\prime}$, and Exo -- covering a wavelength range of approximately 500--1100\,nm. 
This allowed us to test the chromaticity of the transit by measuring the depth individually in each band. 
We find the depths to agree to better than $1.5 \sigma$, and no trend is seen with wavelength.

\subsection{Archival imaging} \label{sec:archival}
\begin{figure*}
    \centering
    \includegraphics[width=.95\textwidth, keepaspectratio]{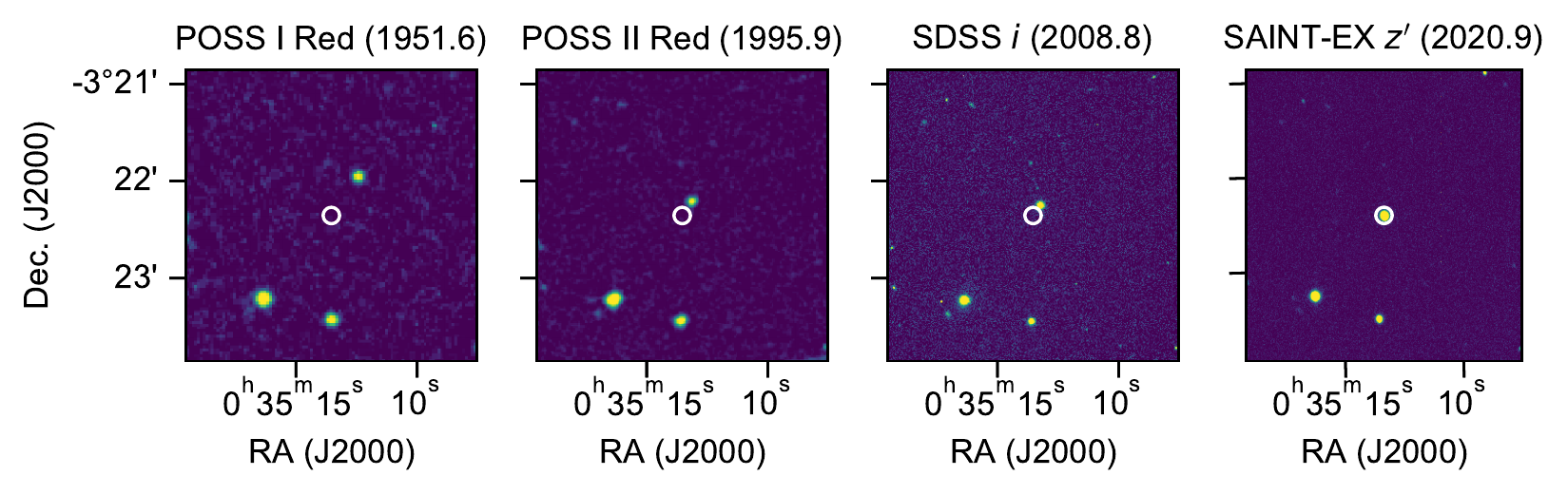}
    \caption{
        Imaging of the current position of TOI-2406, spanning seven decades. 
        Archival images from POSS I, POSS II, and SDSS are shown, along with the median image from the SAINT-EX observations.
        For each image, the bandpass and epoch are noted, and the position of TOI-2406 during the SAINT-EX observations is highlighted.
        The archival images were retrieved using the \texttt{astroquery} \citep{astroquery} interface to \textit{SkyView}\protect\footnote{\url{https://skyview.gsfc.nasa.gov/}}. 
        }
    \label{fig:archival_imaging}
\end{figure*}

We inspected archival images spanning 69 years to constrain the presence of an astrophysical false positive signal at the current position of TOI-2406.
Given its high proper motion \citep[405\,mas/yr,][]{GaiaDR2}, images taken as recently as 2008 are useful for this purpose.
We also note that the nearest neighbour in the Gaia catalogue is at a distance of 65\arcsec and the re-normalised unit weight error (RUWE) is 1.06; that is to say, the target has no close neighbours and its astrometry is explained well by a single-star model \citep{2020MNRAS.496.1922B}.
In total, we considered POSS I/DSS \citep{POSSI, DSS}, POSS II/DSS2 \citep{POSSII}, 2MASS \citep{2006AJ....131.1163S}, and SDSS-III \citep{SDSS-III} imagery, none of which show any point sources at the current position of the target.
Fig.~\ref{fig:archival_imaging} shows a subset of these images.

The SDSS $i$ and $z$ images provide the tightest constraints on background sources. 
The target is offset in these images by 4$\farcs$9 from its current position.
The reported magnitudes of the target are 14.92 and 14.08 in $i$ and $z$, respectively, and no other objects at the current position of the target are detected.
Given the magnitude limits of SDSS DR12\protect\footnote{\url{https://www.sdss.org/dr12/scope/}}, both of these images allow us to rule out sources at the current position of the target that are 6.4\,mag (360 times) fainter than TOI-2406. 
Thus, an undetected background object could contribute no more than 0.3\% of the total flux we observe on-target.
Even the worst-case scenario -- a 100\% drop in brightness on an object at the magnitude limit of the SDSS images -- would not account for the 2\% transit feature we observe.
We conclude that the transit signal originates from the TOI-2406 system and not from another astrophysical source along the same line of sight.

\subsection{False-positive likelihood} \label{sec:fpp}
To fully vet the planet candidate, we utilised two open-source software packages, \texttt{triceratops} \citep{2020ascl.soft02004G, 2021AJ....161...24G} and \texttt{vespa} \citep{2012ApJ...761....6M, 2015ascl.soft03011M}. 
Both of these simulate various false-positive scenarios, allowing a calculation of the likelihood that the transit signal is caused by a planet. 
To statistically validate the planet, we require a false positive probability (FPP) less than 0.01 (1\%), as is typically used in the field.

The \texttt{triceratops} tool was developed to aid in the vetting and validation of TESS objects. The tool uses a Bayesian framework that incorporates prior knowledge of the target star, planet occurrence rates, and stellar multiplicity to calculate the probability that the transit signal is due to a planet transit or another astrophysical source. The resulting FPP quantifies the probability that the transit signal is attributed to something other than a transiting planet. Using the TESS Sector 30 2-minute cadence data, combined with the contrast curve obtained by the Zorro speckle imaging, we obtain an FPP of 0.094, which lies above the threshold for validation. However, the light curve obtained by SAINT-EX provides tighter photometric constraints than the TESS data. We modified the input to \texttt{triceratops} to instead use the SAINT-EX light curve data. The resulting FPP is 0.001 using the light curve alone, and when also including the contrast curve, the FPP is reduced to $3 \times 10^{-11}$. 

We also assessed the candidate with \texttt{vespa}, a similar tool that compares model light curves from distributions of eclipsing binaries (EBs), including hierarchical eclipsing binaries (HEBs) and background eclipsing binaries (BEBs), plus planetary transits. 
The stellar populations first are generated using \texttt{isochrones} \citep{2015ascl.soft03010M}, with inputs of the star's parallax, broadband magnitudes from Table~\ref{table:mags}, and stellar effective temperature and metallicity given in Table~\ref{table:stellar}. 
For inputs into \texttt{vespa}, we used the transit properties from our MCMC analysis (see Sect.~\ref{sec:mcmc}), the star's coordinates and a secondary eclipse threshold of 4.5\,ppt to constrain the false positive scenarios. 
We first evaluated using the 2-minute cadence TESS Sector 30 data and an aperture size of 60\arcsec (3 TESS pixels), finding an FPP of 0.078, again above the threshold required. 
We then used the higher-precision SAINT-EX light curve and an aperture size of 12\arcsec, and found the FPP to be less than $10^{-6}$.

With the \texttt{triceratops} and \texttt{vespa} computed false positive probabilities both much less than one per cent, we consider the candidate signal to be a validated exoplanet, TOI-2406\,b.

\section{Planet searches and detection limits from the TESS photometry} \label{sec:search}

In this section we aim to search for additional planets in the available data and establish detection limits. To search for extra planets, we used our custom pipeline {\fontfamily{pcr}\selectfont  SHERLOCK} \citep{pozuelos:2020,2020A&A...642A..49D} \footnote{{The \fontfamily{pcr}\selectfont  SHERLOCK} (\textbf{S}earching for \textbf{H}ints of \textbf{E}xoplanets f\textbf{R}om \textbf{L}ightcurves \textbf{O}f spa\textbf{C}e-based see\textbf{K}ers) code is fully available on GitHub: \url{https://github.com/franpoz/SHERLOCK}}. The {\fontfamily{pcr}\selectfont  SHERLOCK} package provides the user with easy access to Kepler, K2, and TESS data by searching for and downloading the PDC-SAP flux data from NASA's Mikulski Archive for Space Telescope (MAST). Alternatively, the user may instead provide the data in a \texttt{.csv} file if needed. Then, due to the associated risk of removing transit signals, in particular short and shallow ones, {\fontfamily{pcr}\selectfont  SHERLOCK} applies a multi-detrend approach to the nominal light curve by means of the \texttt{w{\={o}}tan} package \citep{wotan:2019}. Hence, the nominal light curve is detrended a number of times using a bi-weight filter by varying the window size. This strategy allows the user to maximise the signal detection efficiency (SDE) and the S/N of the transit search, which is performed over the nominal light curve, jointly with the new detrended light curves, by means of the \texttt{transit least squares} (TLS) package \citep{tls:2019}. TLS uses an analytical transit model based on the stellar parameters and is optimised for the detection of shallow periodic transits. Once a transit signal is detected with a minimum S/N of 5, {\fontfamily{pcr}\selectfont  SHERLOCK} implements a mask for such a candidate and keeps searching in a new run. This operation is repeated until no more signals with S/N$\geqslant$5 are found in the dataset. 

For Sector 3 we made use of the FFI light curve described in Sect.~\ref{sec:obs}. For Sector 30 we used the 30- and 2-minute cadence light curves provided directly by {\fontfamily{pcr}\selectfont  SHERLOCK}, which utilises the \texttt{ELEANOR} package to access the FFI \citep{eleanor:2019}. In all cases, we recovered the candidate issued by TESS with an orbital period of 3.07\,d in the first run. However, no other signals were found, suggesting that: (1) no other planets are present in the system; (2) if they do exist, they do not transit; or (3) they exist and transit, but the photometric precision of the dataset is not enough to detect them, or they have longer periods than the ones explored with the dataset in hand. If scenario (2) is true, extra planets might be detected by radial velocity follow-up, as discussed in Sect.~\ref{sec:discussion}. 

To evaluate scenario (3), we studied the detection limits of the current dataset by performing injection-and-recovery experiments over the 2-minute cadence PDC-SAP light curve of Sector 30. We made use of the {\fontfamily{pcr}\selectfont  MATRIX ToolKit} \footnote{{The \fontfamily{pcr}\selectfont  MATRIX ToolKit} (\textbf{M}ulti-ph\textbf{A}se \textbf{T}ransits \textbf{R}ecovery from \textbf{I}njected e\textbf{X}oplanets \textbf{T}ool\textbf{K}it) code is fully available on GitHub: \url{https://github.com/martindevora/tkmatrix}}. We explored the $R_{\mathrm{planet}}$--$P_{\mathrm{planet}}$ parameter space in the ranges of 0.7 to 4.0~R$_{\oplus}$ with steps of 0.2~R$_{\oplus}$, and 0.5--10~d with steps of 0.25~d, for a total of 663 different scenarios. {\fontfamily{pcr}\selectfont  MATRIX ToolKit} allows the user to choose a multi-phase approach (i.e. each scenario is explored using a number of different values of T$_{0}$). For simplicity, we assumed the impact parameters and eccentricities of the injected planets were zero. It is worth noting that before to start the search for new candidates we masked the transit times corresponding to TOI-2406~b and we detrended the light curves using a biweight filter with a window-size of 0.18~d, which was found to be the optimal value during the {\fontfamily{pcr}\selectfont  SHERLOCK}'s runs. Moreover, as we injected the synthetic signals in the PDC-SAP light curve, the signals were not affected by the PDC-MAP systematic corrections; therefore, the detection limits should be considered as the most optimistic scenario \citep[see e.g.][]{pozuelos:2020,eisner:2019}.

\begin{figure}
\includegraphics[width=\columnwidth]{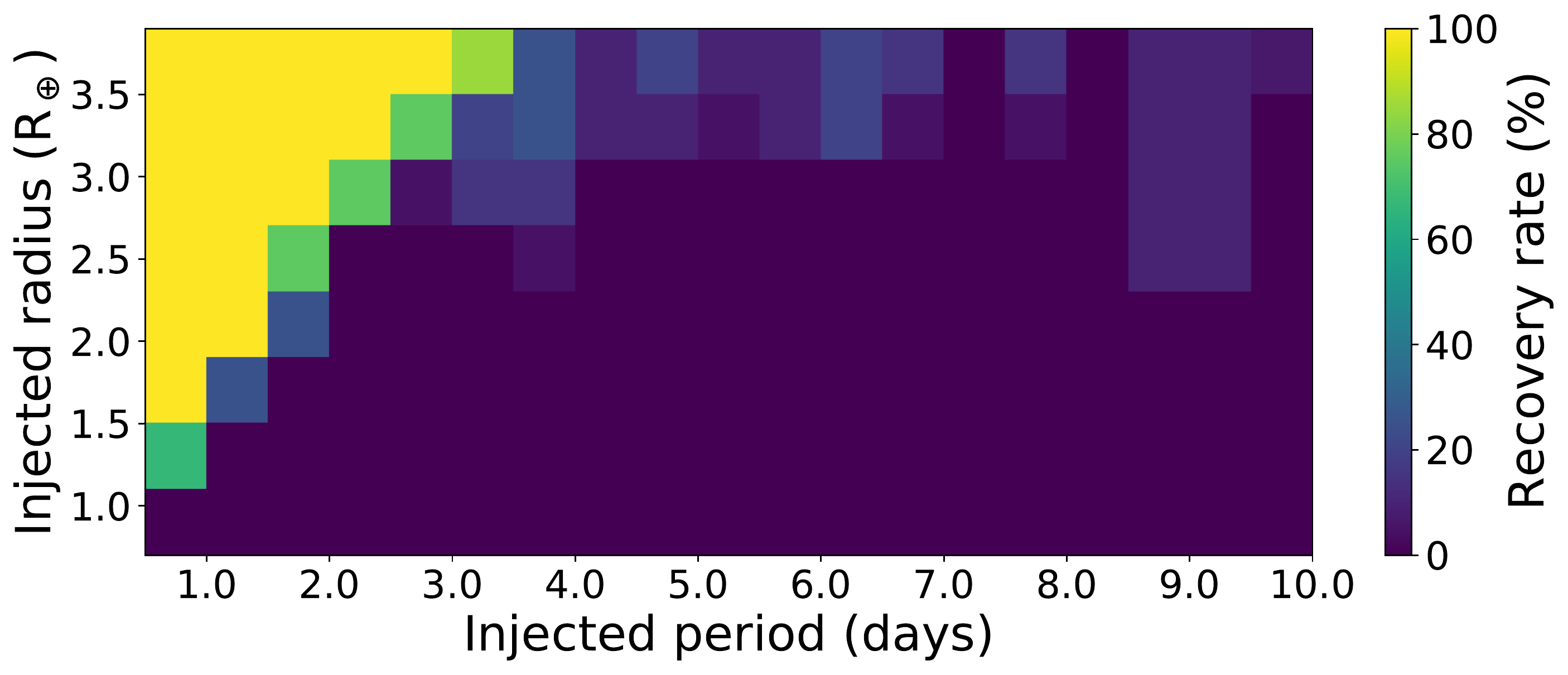}
\caption{Injection-and-recovery tests performed to check the detectability of extra planets in the system. We explored a total of 663 different scenarios and five different phases each for a
total of 3315 simulations. Then, each pixel evaluated about 20 scenarios, that is, 20 light curves with injected planets having different $P_{\mathrm{planet}}$, $R_{\mathrm{planet}}$, and T$_{0}$. Larger recovery rates are presented in yellow and green colours, while lower recovery rates are shown in blue and darker hues. Planets smaller than 1.5~R$_{\oplus}$ would remain undetected for
the full set of periods explored.} \label{recovery}
\end{figure}

In our case, we explored five phases for a total of 3315 different scenarios. A synthetic planet was defined as `recovered' when its epoch was detected with 1~h accuracy and if its period was detected better than 5$\%$. The results are shown in Fig.~\ref{recovery}, which allowed us to reach several conclusions from this test. First, we could hardly detect the presence of planets of any size in the range explored with orbital periods $\gtrsim$4~d, where we obtained a recovery rate of $\lesssim$20$\%$. Second, for orbital periods $\leq$4~d we could detect smaller planets, down to a minimum size of $\sim$1.5~R$_{\oplus}$ for orbital periods $\leq$1~d. Hence, we may rule out the presence of such planets; if they existed and transited, they would be easy to detect, with recovery rates ranging from 80 to 100$\%$. Finally, planets smaller than 1.5~R$_{\oplus}$ would remain undetected for the full set of periods explored.
 
%-----------------------------------------------------------------

\section{Transit analysis} \label{sec:mcmc}

\subsection{MCMC procedure}
All the light curve data detailed in Sect.~\ref{sec:obs} were fit simultaneously via an MCMC procedure. 
To do this, we utilised the \texttt{PyTransit} \citep{Parviainen2015} implementation of the \citet{2002ApJ...580L.171M} quadratic limb-darkening transit model, and the \texttt{emcee} \citep{2013PASP..125..306F} implementation of the affine-invariant ensemble sampler \citep{2010CAMCS...5...65G}. 

We fitted eight parameters of the system: the planet-to-star radius ratio ($\mathrm{R_p}$/$\mathrm{R_*}$), transit epoch ($\mathrm{T_0}$), orbital period (P), impact parameter (b), eccentricity (e), longitude of periastron ($\mathrm{\omega}$) and the stellar radius ($\mathrm{R_*}$) and mass ($\mathrm{M_*}$). 
The planet's semi-major axis in stellar radii ($\mathrm{a / R_*}$) was computed from the stellar density (from $\mathrm{R_*}$, $\mathrm{M_*}$) and orbital period, using Kepler's third law. 
We also fitted the quadratic limb-darkening coefficients (LDCs), $u_1$ and $u_2$, for each of the five photometric bands covered by the transit data (TESS, V, $\mathrm{i^\prime}$, $\mathrm{z^\prime}$, Exo). 
Gaussian priors for the LDCs were computed using \texttt{PyLDTk} \citep{Parviainen-ldtk}, which uses the stellar spectrum model library of \citet{Husser2013}. 
We increased the widths of the computed priors by a factor of five to account for the model-dependent uncertainties. 
For each dataset, we included the out-of-transit baseline as a free parameter with a Gaussian prior centred on unity with a width of 0.01. 
Detrending vectors for airmass and the full width at half maximum were also fit for the ground-based data. 
An extra vector for sky background was also included for the EDEN/VATT data, because observations started during twilight. 
Each vector was generated by multiplying the telescope-recorded property by a scale factor, which was fitted in the MCMC. 
Each scale factor had a normal prior centred on zero, with a width of 0.1. 

\subsection{Results}
The MCMC detrended and transit modelled light curves are shown in Fig.~\ref{fig:transits}. 
The posterior distributions for the eight fitted transit properties can be found in Appendix~\ref{app:corner}. 
The fitted transit properties are given in Table~\ref{table:planet} with their 1$\sigma$ confidence levels. 
Most notably, we find the planet to have a radius of $2.94 \pm 0.17\,\mathrm{R_\oplus}$ and an eccentricity of $0.26_{-0.12}^{+0.27}$, which leads to longer transits than compared to a circular orbit. 
We also predict some properties based on the estimated planet mass. 
We compared planetary masses predicted by the mass-radius relationships of \citet{2017ApJ...834...17C}, \citet{2018ApJ...869....5N} and \citet{2019ApJ...882...38K}. 
We find the mass posteriors of both \citet{2018ApJ...869....5N} and \citet{2019ApJ...882...38K} are considerably lower than for \citet{2017ApJ...834...17C}, and extend down to unphysically low masses. 
We therefore elect to use the \citet{2017ApJ...834...17C} mass estimate in this work, but reason it could be somewhat over-estimated and so could be seen as more of an upper limit.

\begin{figure*}[htbp!]
\centering
\includegraphics[width=17cm]{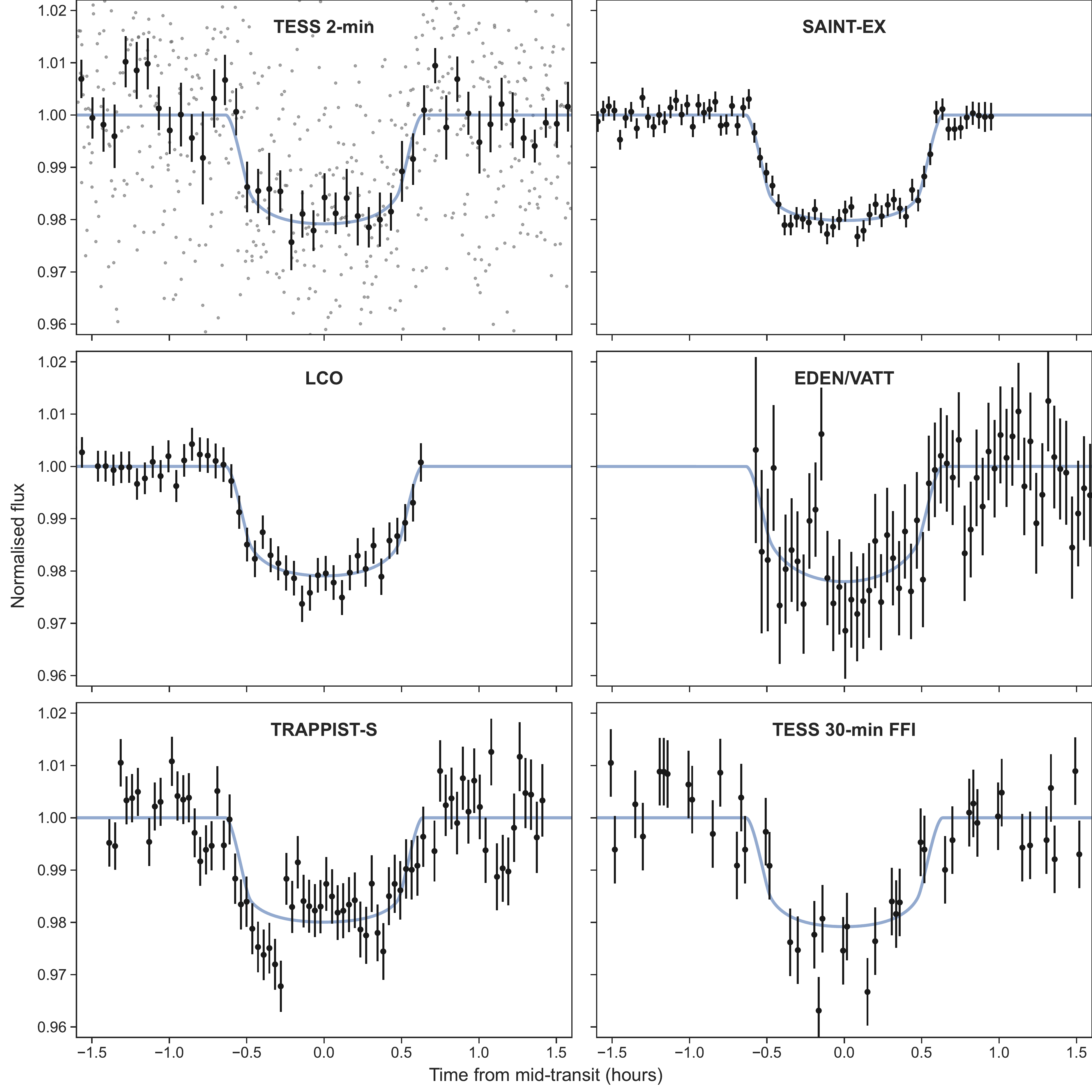}
\caption{Phase-folded transit light curves of TOI-2406\,b. From left to right and from top to bottom: TESS Sector 30 2-minute-cadence, SAINT-EX z$^\prime$, LCO i$^\prime$, EDEN/VATT V, TRAPPIST-South Exo, and TESS 30-minute FFI. 
Each light curve has been corrected for the median trend and baseline from the MCMC fit. 
The median transit model is also over-plotted. 
The TESS 2-minute data (grey points) have been binned by a factor of 15 to produce the black data points, with error bars corresponding to the scatter within each bin.}
\label{fig:transits}
\end{figure*}

\begin{table}
\caption{Median properties with 1$\sigma$ confidence levels, from the transit analysis. 
$^\dagger$Calculated with a Bond albedo of zero. 
} 
\label{table:planet}
\setlength{\tabcolsep}{12pt}
\centering
\begin{tabular}{l c}
\hline\hline
Property & Value \\[0.1cm]
\hline
Fitted parameters: & \\[0.1cm]
$\mathrm{T_0}$ (BJD$- 2450000$) & $9115.97547 \pm 0.00027$ \\[0.1cm]
P (d) & $3.0766896 \pm 6.5 \times 10^{-6}$ \\[0.1cm]
$\mathrm{R_p}$/$\mathrm{R_*}$ & $0.1322 \pm 0.0020$ \\[0.1cm]
b ($\mathrm{R_*}$) & $0.16_{-0.11}^{+0.15}$ \\[0.1cm]
$\mathrm{\sqrt{e}\cos\omega}$ & $0.06_{-0.55}^{+0.45}$ \\[0.1cm]
$\mathrm{\sqrt{e}\sin\omega}$ & $-0.358_{-0.095}^{+0.111}$ \\[0.1cm]
$\mathrm{R_*}$ ($\mathrm{R_\odot}$) & $0.204 \pm 0.011$ \\[0.1cm]
$\mathrm{M_*}$ ($\mathrm{M_\odot}$) & $0.162 \pm 0.008$ \\[0.4cm]

Limb-darkening: & \\[0.1cm]
$\mathrm{u_1}$ TESS & $0.313 \pm 0.059$ \\[0.1cm]
$\mathrm{u_2}$ TESS & $0.39 \pm 0.11$ \\[0.1cm]
$\mathrm{u_1}$ $\mathrm{z^\prime}$ & $0.240 \pm 0.045$ \\[0.1cm]
$\mathrm{u_2}$ $\mathrm{z^\prime}$ & $0.354 \pm 0.088$ \\[0.1cm]
$\mathrm{u_1}$ $\mathrm{i^\prime}$ & $0.337 \pm 0.066$ \\[0.1cm]
$\mathrm{u_2}$ $\mathrm{i^\prime}$ & $0.37 \pm 0.11$ \\[0.1cm]
$\mathrm{u_1}$ V & $0.56 \pm 0.13$ \\[0.1cm]
$\mathrm{u_2}$ V & $0.21 \pm 0.17$ \\[0.1cm]
$\mathrm{u_1}$ Exo & $0.268 \pm 0.064$ \\[0.1cm]
$\mathrm{u_2}$ Exo & $0.25 \pm 0.12$ \\[0.4cm]

Derived parameters: & \\[0.1cm]
$\mathrm{R_p}$ ($\mathrm{R_\oplus}$) & $2.94_{-0.16}^{+0.17}$ \\[0.1cm]
$\mathrm{\rho_*}$ & $26.9_{-4.4}^{+4.9}$ \\[0.1cm]
$\mathrm{a / R_*}$ & $24.0_{-1.1}^{+1.0}$ \\[0.1cm]
a (AU) & $0.0228 \pm 0.0016$ \\[0.1cm]
i ($^\circ$) & $89.63_{-0.35}^{+0.27}$ \\[0.1cm]
e & $0.26_{-0.12}^{+0.27}$ \\[0.1cm]
$\mathrm{\omega}$ ($^\circ$) & $279_{-63}^{+47}$ \\[0.1cm]
$\mathrm{S_p}$ ($\mathrm{S_\oplus}$) & $6.55_{-0.80}^{+0.94}$ \\[0.1cm]
$\mathrm{T_{eq}}^\dagger$ (K) & $447 \pm 15$ \\[0.4cm]

Predicted parameters: & \\[0.1cm]
$\mathrm{M_p}$ ($\mathrm{M_\oplus}$) & $9.1_{-4.0}^{+7.1}$ \\[0.1cm]
K (m/s) & $14.9_{-6.6}^{+12.0}$ \\[0.1cm]
TSM & $115_{-50}^{+87}$ \\[0.1cm]
ESM & $4.12_{-0.58}^{+0.67}$ \\[0.1cm]
\hline
\end{tabular}
\end{table}

\subsection{Eccentricity} \label{sec:mcmc-ecc}
\begin{figure}
\includegraphics[width=\columnwidth]{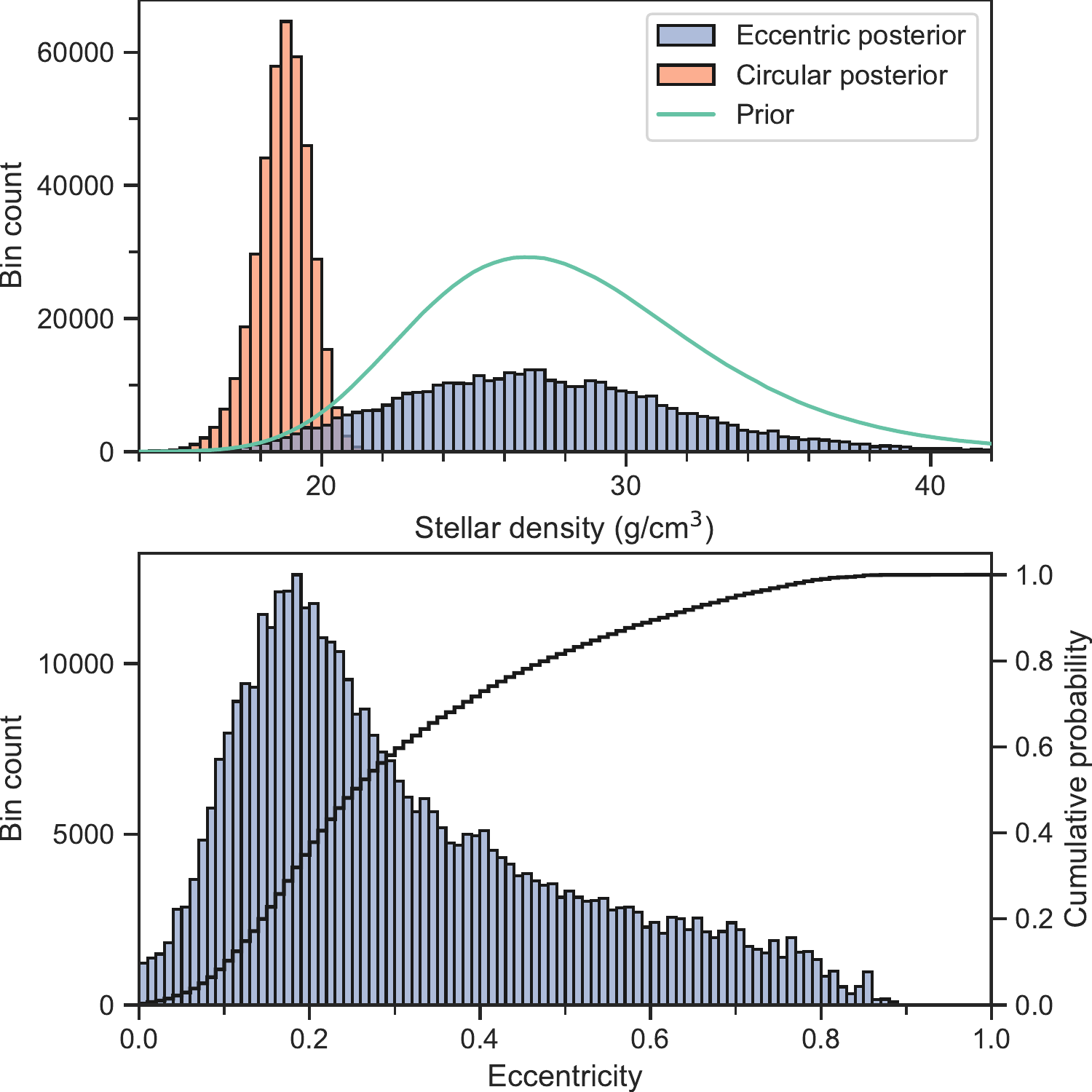}
\caption{Posterior distributions from the MCMC fitting procedure. Top: Stellar density posterior distributions for both the eccentric and circular orbit cases. 
The prior distribution is over-plotted (arbitrarily scaled), calculated from the stellar mass and radius in Table~\ref{table:stellar}. 
Bottom: Eccentricity posterior distribution for the eccentric orbit case. The cumulative probability curve is over-plotted, showing very little agreement with a circular orbit.} \label{fig:density-ecc}
\end{figure}

Our transit model fit heavily favours a non-zero eccentricity (see Fig.~\ref{fig:density-ecc}) and a longitude of periastron around $270^\circ$ ($\cos\omega = 0$). 
This is due to the prior on the stellar density (through the stellar radius and mass) and the photometry exhibiting a longer transit duration than expected for a circular orbit. 
Given the old age of the host star, one might expect eccentricity dampening to have produced a near-circular orbit by now. 
We discuss this is detail in Sect.~\ref{orbi_ex}.
To test the case of a circular orbit, we proceeded with another MCMC analysis, similar to the one described above, with the exception of fixing the eccentricity at zero. 
While a good fit was obtained, the values of the fitted stellar radius ($0.227 \pm 0.005$ $\mathrm{R_\odot}$) and mass ($0.156 \pm 0.008$ $\mathrm{M_\odot}$) differ from the priors by 2$\sigma$ and 1$\sigma$, respectively. 
As shown in Fig.~\ref{fig:density-ecc}, this corresponds to a much lower stellar density compared to the eccentric case, which is at odds with our inferred stellar properties. 

To confirm our analysis was reliable, we also used \texttt{EXOFASTv2} \citep{Eastman2013,2019arXiv190709480E} to  simultaneously fit (a) the observed TESS and SAINT-EX light curves with a transit model, (b) the broadband photometry and the \textit{Gaia} EDR3 parallax (assuming zero A$_{\rm V}$ extinction) with an SED based on MIST atmospheres, and (c) constraints on the stellar mass and radius with MIST stellar evolution models. 
Thus, our \texttt{EXOFAST} analysis allows us to explore the circular and eccentric cases, while enforcing the astrophysical constraint of a realistic stellar radius for a low-mass M dwarf.
We used a Gaussian prior on the metallicity and the parallax, and because TOI-2406 is close to the 0.1~M$_\odot$ limit of the MIST stellar models, we used a Gaussian prior on the mass with the values reported in Sect.~\ref{sec:stellar}, which are based on stellar models that more appropriate for an M4 dwarf. 
We used uniform priors for the stellar radius and for the effective temperature. 
For the eccentric case, we find very similar results, with a non-zero eccentricity preferred (e=$0.35_{-0.15}^{+0.33}$), while the stellar and planetary and parameters are consistent within their 1$\sigma$ uncertainties with those in Table~\ref{table:planet}. 
Forcing the orbit to be circular and, given the stellar radius constraints from the SED and stellar models, \texttt{EXOFAST} is not able to fit the SAINT-EX photometry. 
In other words, we do not find a solution with a circular orbit that is consistent with a physically realistic radius of the star.  

As detailed in Sect.~\ref{sec:sedstellar}, there is good agreement between empirical estimates of the stellar mass and radius and our stellar models. 
We are therefore confident in the accuracy of our parameter estimates, and, consequently, also in our stellar density estimate.
Thus, given the above, we consider that a circular orbit is not likely for TOI-2406\,b. We present our preferred best fit and resulting stellar and planet properties for an eccentric orbit in Table~\ref{table:planet}. For completeness, we include the circular orbit solution in Appendix~\ref{app:circular}, but we emphasise that a circular orbit is not preferred from the current data and analysis, and tends to an unphysical stellar radius.

%-----------------------------------------------------------------

\section{Discussion} \label{sec:discussion}

\subsection{Formation}
The formation of close-in planets is commonly explained by core accretion and migration \citep[e.g.][]{2013A&A...558A.109A, 2019A&A...627A..83L}. Both planetesimal-based \citep{2004ApJ...616..567I,2012A&A...541A..97M,2020arXiv200705562E} and pebble-based \citep{2018A&A...619A.174B,2019A&A...623A..88B} models are able to reproduce the observed metallicity correlation for giant planets \citep{1997MNRAS.285..403G,2004A&A...415.1153S}. For lower planetary masses, the general metallicity trend weakens \citep{2011arXiv1109.2497M,2012Natur.486..375B} but remains positive for planets at $P<10$ days  \citep{2016AJ....152..187M,2018AJ....155...89P}. In the following, we briefly put the detection of a large sub-Neptune around a low-metallicity host star into the perspective of current planet formation models.

Planetesimal-based models \citep[e.g.][]{2020arXiv200705561E} assume the bulk of the solid material to coagulate into planetesimals at an early time. Planetesimal formation models predict a significantly steeper planetesimal surface density due to radial drift of pebbles \citep{2017A&A...608A..92D,2019ApJ...874...36L}. This process can increase the amount of solids available in the inner regions of the disk, but the total solid mass in the system remains correlated to the dust to gas ratio of the disk. As the growth of inward-migrating planets -- in this planetesimal accretion scenario -- is limited by the solid reservoir interior to its starting location \citep{2008ApJ...673..487I}, a statistical correlation with the initial dust-to-gas ratio of the disk is predicted by these models. The observed stellar metallicities are taken as a proxy to prescribe the total dust-to-gas ratio of the disk.

While this is mostly relevant for large planetary masses around solar-type stars, it becomes important for low-mass planets around lower stellar masses. Recent ALMA measurements showed that the disk mass is related to the stellar mass with a steeper-than-linear trend \citep{2016ApJ...831..125P,2017AJ....153..240A}. Therefore, much less material might be present around low-mass stars.

The recent planetesimal-based simulations of \citet{Burn2021} linearly scale Class I disk-mass estimates \citep{2018ApJS..238...19T} to stellar masses of $0.1\,M_{\odot}$. The reason for the linear scaling is due to an identified evolutionary trend in the Class II measurements \citep{2016ApJ...831..125P,2017AJ....153..240A} and can therefore be interpolated back in time to get an 'initial' relation. The \citet{Burn2021} work is based on the updated Bern model of planet formation \citep{2020arXiv200705561E} and one of the recent theoretical works addressing low-mass stars and making use of the improved observational disk constraints. The model is well suited for comparison to observations by consistently evolving planetary atmospheres over Gyr timescales in order to calculate planetary radii. The results are in general agreement to studies more focused on individual aspects (e.g. \citealp{2019A&A...631A...7C,2019A&A...627A.149S} for the TRAPPIST-1 system, or \citealp{2020MNRAS.491.1998M} who do not vary the stellar metallicity or dust-to-gas ratio and can therefore not be used for our purposes). 

\begin{figure}
    \centering
    \includegraphics[width=\linewidth]{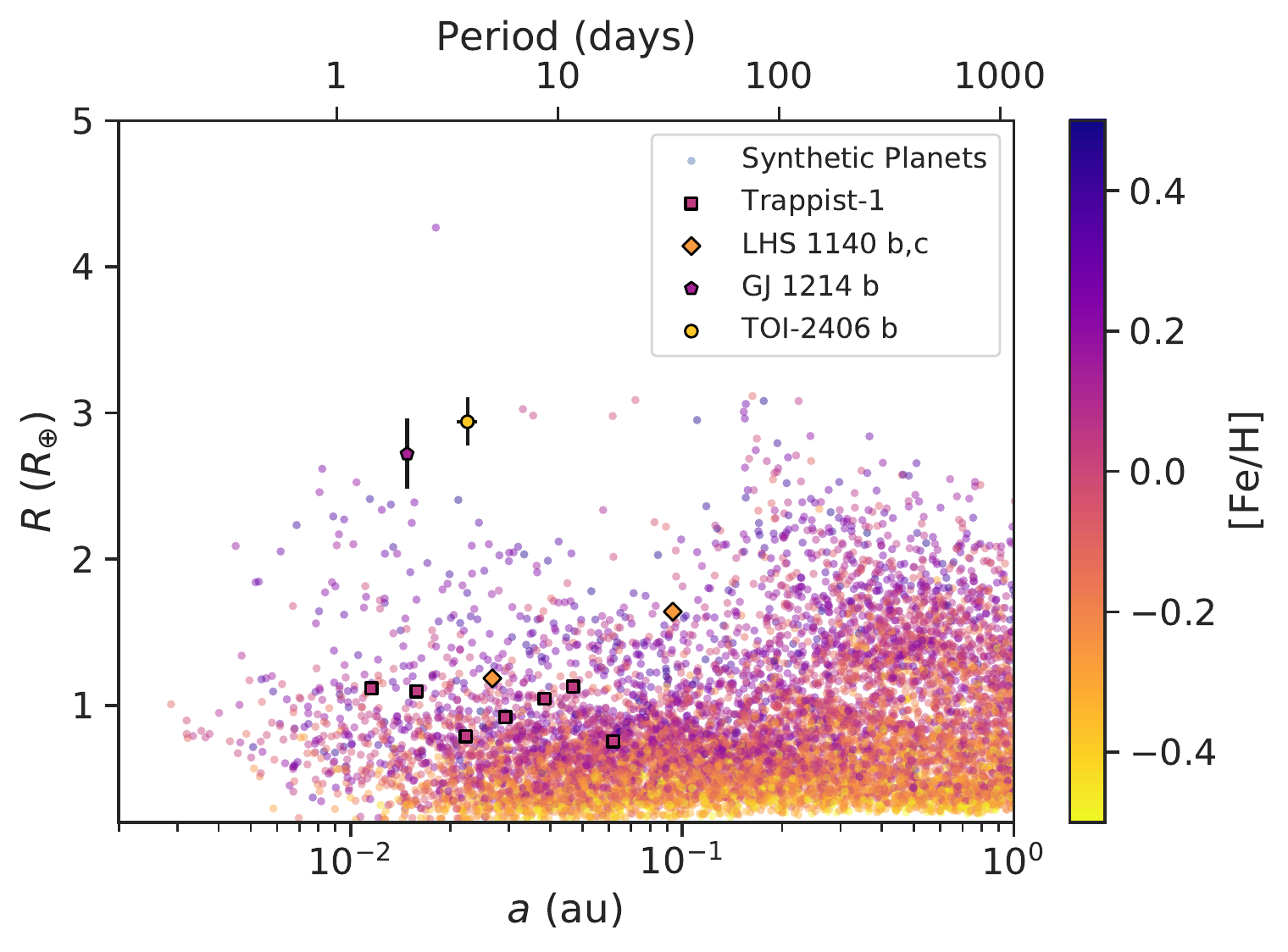}
    \caption{Semi-major axis versus radius distribution of synthetic and observed planets. To put TOI-2406\,b into perspective, the population of predicted core-accretion planets from the $0.1\,M_{\odot}$ case in \citet{Burn2021} is shown. The metallicity of the host stars is colour-coded and used in the theoretical calculations as a proxy for the initial planetesimal-to-gas-mass ratio. The metallicity and planetary radii for TRAPPIST-1 are taken from \citet{2018MNRAS.475.3577D} and \citet{2021PSJ.....2....1A}, respectively; those for LHS 1140 from \citet{2019AJ....157...32M} and \citet{2020A&A...642A.121L}; and those for GJ 1214 from \citet{2013A&A...551A..48A}. Those systems are in better agreement with the theoretical calculations than TOI-2406\,b.}
    \label{fig:synthetic}
\end{figure}

In the population synthesis results of \citet{Burn2021} for a 0.1$\,M_{\odot}$ star, which are shown in Fig. \ref{fig:synthetic}, a strong dependence on the sampled metallicity is present. In these planetesimal-accretion simulations, the conditions around low-metallicity hosts never allows for growth of planets to sizes comparable to TOI-2406\,b. Instead, only in a few high-metallicity disks -- out of the 1000 cases simulated -- some similar-sized planets form.

Reproducing TOI-2406\,b in a planetesimal-based, core-accretion model is very challenging.
Given the steeper-than linear ALMA measurements, it is unrealistic to further increase total disk masses. Another pathway to increase growth is a reduction in migration timescales. However, this is only efficient where planets grew to significant mass in the early, in situ growth phase. Given the already accretion-favourable parameter choices for planetesimal sizes and dust opacities in the simulations, we conclude that the detection of TOI-2406\,b poses a significant challenge for core-accretion models based on planetesimal accretion.

If the main solid accretion channel is dominated by pebbles instead of planetesimals \citep{2010A&A...520A..43O}, growth is limited by the pebble isolation mass, which scales linearly with the stellar mass \citep{2019A&A...632A...7L}. Therefore, the limit does not depend (to first order) on metallicity. However, there is an indirect dependence in a cooling protoplanetary disk: longer pebble accretion timescales in low-metallicity disks lead to planets reaching the pebble isolation mass at later times. Therefore, the temperature has further decreased, and the scale height is reduced \citep{2019A&A...623A..88B}. This is also why the environment of the star can influence the metallicity correlation \citep{2018MNRAS.474..886N}.

The analysis in the pebble accretion scenario for low-mass star by \citet{2019A&A...632A...7L} shows some potential to form Earth-mass planets around late M dwarfs in rare cases with only a weak dependence on stellar mass. However, \citet{2019A&A...632A...7L} assumed the presence of millimetre-sized pebbles independent of metallicity. It remains to be explored if grains can coagulate on short-enough timescales to build up the required flux of drifting pebbles at low metallicities \citep{2016SSRv..205...41B} and if seed planetesimals can grow \citep[e.g. by streaming instability]{2009ApJ...704L..75J}. In principle, dust growth should be significantly slower in low-metallicity environments. However, growth and therefore pebble accretion timescales are also sensitive to turbulence, fragmentation thresholds, and disk sizes \citep{2021A&A...647A..15D}.

There is a strong tension between planetesimal accretion predictions and TOI-2406\,b. In contrast, its presence is more naturally produced in the pebble accretion scenario, which is therefore the favoured mode of accretion for this planet. 
However, open questions about the required pebble and planetesimal growth remain. 
Also, we cannot rule out formation via gravitational instability, where the planet could form far out in the disk and then migrate and downsize \citep{2010MNRAS.408L..36N,2018MNRAS.474.5036F}.
TOI-2406 therefore presents a benchmark case for future investigations: a large sub-Neptune planet around a significantly metal-poor M dwarf.

\subsection{Orbital excitation}\label{orbi_ex}
Given the age of the system is expected to be ${\sim} 11$\,Gyr, we would not expect the planet to retain a non-zero eccentricity. 
Following \citet{1966Icar....5..375G}, using a tidal quality factor (Q$^\prime$) of $10^5$, similar to the value expected for Neptune (0.9--33$ \times 10^4$, \citealt{1966Icar....5..375G,1992Icar...99..390B,2008Icar..193..267Z}), and the predicted planetary mass, we find an eccentricity dampening timescale of 0.25\,Gyr. 
This is much shorter than the expected age of the system, and therefore we would expect the planet's orbit to have been circularised by tidal dissipation. 
However, we find an eccentricity of $\sim$0.2 is needed to fit the transit (see Sect.~\ref{sec:mcmc-ecc}). 
A value for Q$^\prime$ of $4.4 \times 10^6$ is needed for the timescale to approach the expected age of the system, which is much higher than estimated for Neptune. 

TOI-2406 appears to be the fourth planetary system to host a single (sub-) Neptune on a short (P<10\,d), eccentric orbit; the others being, GJ~436 \citep{2004ApJ...617..580B}, GJ~674 \citep{2007A&A...474..293B}, and TOI-269 \citep{2021arXiv210414782C}. 
The GJ~436 system has been particularly well studied, which consists of a 0.4\,$\mathrm{M_\odot}$ M dwarf, with a Neptune on an eccentric ($e = 0.162 \pm 0.004$; \citealt{2014A&A...572A..73L}) 2.6\,d orbit. 
For this system, another solution has been proposed where the planet's eccentricity is caused by interaction with another planet or a bound star \citep{2007PASP..119...90M,2008ApJ...677L..59R,2012A&A...545A..88B}. 
However, searches for such companions have not been rewarding \citep{2009IAUS..253..149R,2010PASP..122.1341B,2014A&A...572A..73L}. 
One further possibility is that a recent encounter with a nearby star may have perturbed the orbit, and the circularisation will complete in the next few hundred megayears. 
Here we highlight a possible trend in these systems, noting that the host stars are all mid-M dwarfs, pointing to a likely common mechanism allowing eccentricity to be raised and remain high. 

TOI-2406 represents the most extreme difference between age of the system and its tidal circularisation timescale. 
For TOI-2406\,b, it is tempting to explain this disparity by a bound companion to the star. 
This companion could cause an eccentricity excitation via direct dynamical interaction or through Lidov-Kozai cycles \citep{1962P&SS....9..719L,1962AJ.....67..591K,2007ApJ...670..820W}. 
An object more massive than an $\sim$M8 star is ruled out beyond roughly 6\,AU by speckle imaging, and hotter stars within this distance would be seen in the near-infrared spectrum. 
Therefore, the mass of any potential companion is limited to lower-mass objects, such as a brown dwarf or a second planet in the system. 
Such objects could be detected by a future radial velocity programme.

\subsection{Potential for radial velocity observations}
Deriving the mass of TOI-2406\,b would not only give us another validation of the planetary nature but would also allow us to derive the full orbital parameters, such as the argument of periastron ($\mathrm{\omega}$), as well as better constrain the eccentricity. Better constraining the eccentricity will in turn help us to put constraints on the dynamical history of this system (see Sect. \ref{orbi_ex}).

The mass we have estimated for TOI-2406\,b, combined with the derived orbital parameters, indicates a semi-amplitude of $14.9_{-6.6}^{+12.0}$ m/s. Given the faintness of the star, the detection of this shallow signal is challenging even for state-of-the-art instrumentation and requires a great deal of observing time. For stabilised optical spectrographs like ESPRESSO \citep{pepe10}, fed by the 8 m unit telescope of the VLT, we expect a radial-velocity (RV) precision of about 40\,m/s for a single measurement with a 15\,min exposure. Thus, 80 to 100 measurements would be needed for a 3$\sigma$ detection of the RV signal.

Stabilised high-resolution infrared spectrographs have become operational in the past years that  make it possible to investigate such very red stars more efficiently. For example, using the same telescope aperture and same exposure time as for ESPRESSO, but instead the newly commissioned CRIRES$^{+}$ \citep{dorn14} spectrograph (which will become available in October 2021), would lead to one order of magnitude better precision. Assuming that the instrumental systematics are well understood \citep{Figueira10}, a 3$\sigma$ detection of the RV signal would be possible with only 12 measurements. The combined measurements would allow the stellar parameters as well as the $v \sin i$ of TOI-2406 to be further constrained. Furthermore, any deviation from the expected RV signal would allow one to draw hints on the presence of possible further bodies orbiting TOI-2406.

Given this further prospect, the direct mass determination of TOI-2406\,b will soon become feasible and allow us to better constrain its composition and atmospheric parameters for future transmission spectroscopy.

\subsection{Potential for transmission spectroscopy}

\begin{figure}[t!]
    \centering
    \includegraphics[width=\columnwidth]{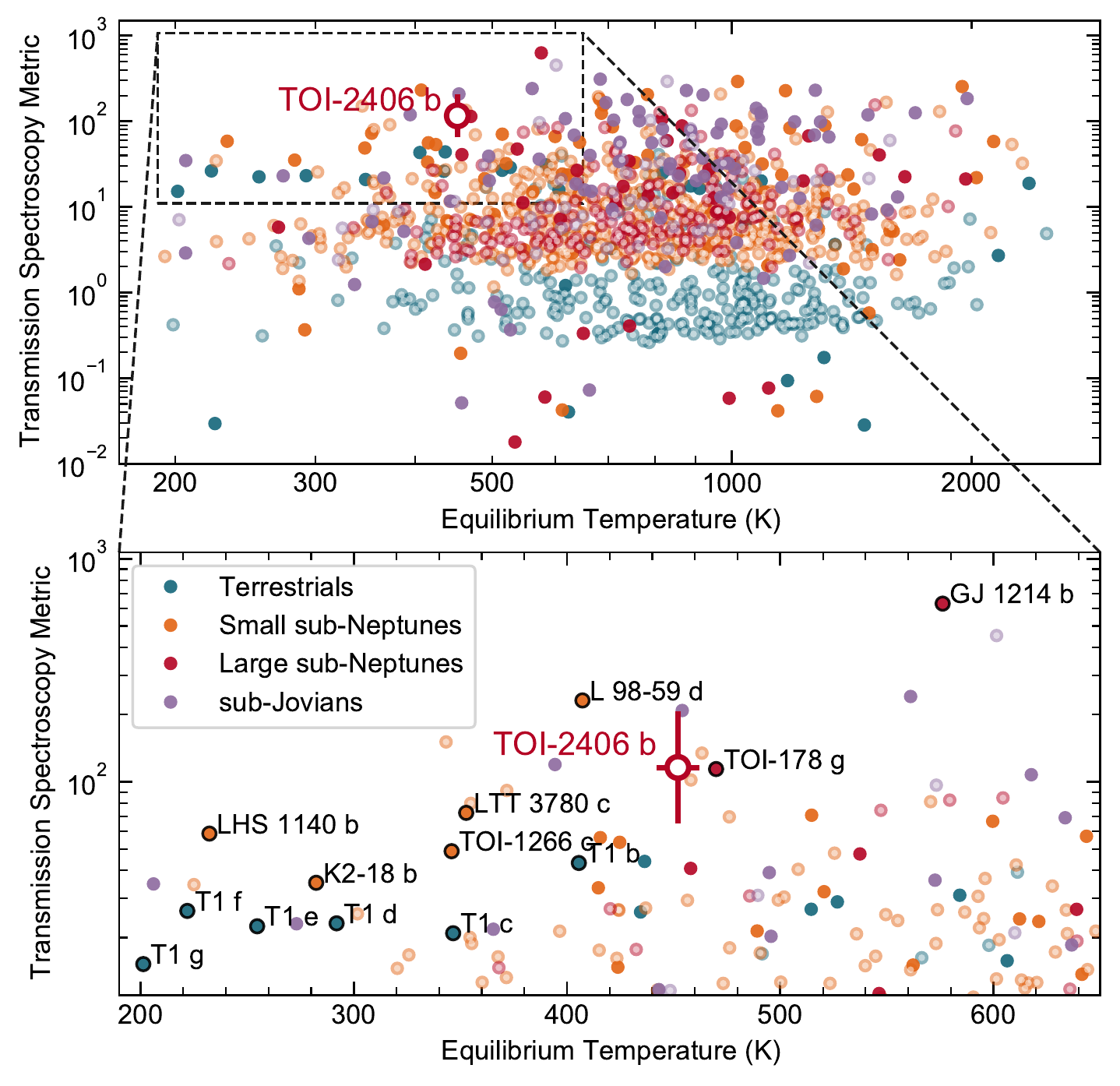}
    \caption{
    Suitability of TOI-2406\,b for transmission spectroscopy studies.
    The top panel shows the complete sample of known transiting exoplanets, while the bottom panel focuses on the coolest ones amenable to study.
    TOI-2406\,b, highlighted in both panels, is among the most accessible cool planets for atmospheric studies.
    Planets are colour-coded by the radius bins of \citet{Kempton2018}.
    Filled circles indicate planets with mass measurements, while outlined circles indicate those with masses estimated from empirical mass-radius relations.
    Some highly accessible planets smaller than Neptune are named and highlighted with black outlines in the bottom panel.
    }
    \label{fig:tsm}
\end{figure}

One of the primary allures of planets transiting small stars is the potential to study their atmospheres in detail. 
Relatively cool planets like TOI-2406\,b are most amenable to studies via transmission spectroscopy \citep[e.g.][]{Lustig-Yaeger2019}. 
Thus, we consider here atmospheric follow-up of TOI-2406\,b via this technique. 

We assessed the potential for studies of TOI-2406\,b in transmission using the transmission spectroscopy metric \citep[TSM;][]{Kempton2018}. 
We calculated the TSM for all transiting exoplanets in the NASA Exoplanet Archive\footnote{Accessed 24 Feb 2021.} with a reported stellar radius and effective temperature, and a planetary radius and semi-major axis. 
Following the definition of the TSM, we estimated planetary masses for those without measured ones using the empirical mass-radius relation \citep{2017ApJ...834...17C, Louie2018}. 
Using the planet-size bins of \citet{Kempton2018}, this translated to samples of 296, 611, 202, and 134 measurements for terrestrials ($R_p < 1.5\,R_\oplus$), small sub-Neptunes ($1.5 < R_p < 2.75\,R_\oplus$), large sub-Neptunes ($2.75 < R_p < 4.0\,R_\oplus$), and sub-Jovians ($4.0 < R_p < 10\,R_\oplus$), respectively. 
Using the parameters derived here (\autoref{table:planet}), we also calculated the TSM for TOI-2406\,b, a planet that we note falls into the large sub-Neptune size bin at $2\sigma$ confidence. 

As seen in Fig.~\ref{fig:tsm}, we find that the TSM of TOI-2406\,b is higher than those of all but 3.0\% of the 1243 planets we surveyed. 
For the large sub-Neptunes, it is higher than all but 2.0\% of the sample. 
In fact, only four large sub-Neptunes have higher TSMs: GJ\,1214\,b \citep{Charbonneau2009}, K2-266\,b \citep{Rodriguez2018}, HD\,191939\,b \citep{Badenas-Agusti2020}, and TOI-1130\,b \citep{Huang2020}. 
These have equilibrium temperatures, assuming zero Bond albedo and full day-night heat redistribution (following the definition of the TSM), of 580--1510 K -- significantly hotter than the $447 \pm 15$\,K equilibrium temperature of TOI-2406\,b\footnote{Using a Bond albedo of 0.3, which is observed for Earth, Uranus, and Neptune and might be more realistic for large sub-Neptunes generally, these quoted equilibrium temperatures decrease by 10\%.}.

While the final say in the suitability of TOI-2406\,b awaits a mass measurement, it is likely a target uniquely suited for studies of relatively cool atmospheres in the large sub-Neptune regime. 
\citet{Crossfield2017} note that the amplitude of atmospheric features in transmission for warm Neptune-sized planets correlate with either equilibrium temperatures or bulk H/He mass fractions. 
They argue that these correlations could point to either hazier atmospheres at lower temperatures or more metal-rich atmospheres for smaller planets (or both). 
Compared to the warm Neptune sample analysed in that work, TOI-2406\,b is both relatively cool and small. 
Studies of its atmosphere in transmission with HST and JWST could thus be useful for further testing both trends suggested by the current sample of Neptune-sized planets.

%-----------------------------------------------------------------

\section{Conclusions} \label{sec:conclusion}

We have presented the discovery and initial analysis of the planet TOI-2406\,b. 
The system consists of a large sub-Neptune orbiting a low-mass member of the thick disk. 
It is a challenge to the core accretion model of planet formation, where simulations struggle to produce large planets around metal-poor, late-type stars. 
This is particularly problematic for planetesimal-based models: They may point towards a pebble accretion formation for TOI-2406~b, which is more favourable. 
The planet also has a significant non-zero eccentricity at an age far beyond the estimated circularisation timescale. 
Furthermore, the planet is expected to be a good candidate for transmission spectroscopy in the warm Neptune regime with JWST. 
However, a stronger prediction of the expected S/N awaits a direct mass detection from radial velocity observations, which are possible with the latest infrared spectrographs at 10-metre-class telescopes.

%-----------------------------------------------------------------

\begin{acknowledgements}

We thank the anonymous referee for their corrections and help in improving the paper. 

We warmly thank the entire technical staff of the Observatorio Astron\'omico Nacional at San Pedro M\'artir in M\'exico for their unfailing support to SAINT-EX operations, namely:
E.~Cadena, T.~Calvario, E.~Colorado, B.~Garc\'ia, G.~Guisa, A.~Franco, L.~Figueroa, B.~Hern\'andez, J.~Herrera, E.~L\'opez, E.~Lugo, B.~Mart\'inez, J.M.~Nu\~nez, J.L.~Ochoa, M.~Pereyra,, F.~Quiroz, T.~Verdugo, I. Zavala. 

B.V.R. thanks the Heising-Simons Foundation for support.
Y.G.M.C acknowledges support from UNAM-PAPIIT IG-101321.
B.-O. D. acknowledges support from the Swiss National Science Foundation (PP00P2-163967 and PP00P2-190080).
R.B. acknowledges the support from the Swiss National Science Foundation under grant P2BEP2\_195285.
M.N.G. acknowledges support from MIT's Kavli Institute as a Juan Carlos Torres Fellow.
A.H.M.J.T acknowledges funding from the European Research Council (ERC) under the European Union's Horizon 2020 research and innovation programme (grant agreement n$^\circ$ 803193/BEBOP), from the MERAC foundation, and from the Science and Technology Facilities Council (STFC; grant n$^\circ$ ST/S00193X/1).
T.D. acknowledges support from MIT's Kavli Institute as a Kavli postdoctoral fellow

Part of this work received support from the National Centre for Competence in Research PlanetS, supported by the Swiss National Science Foundation (SNSF).

The research leading to these results has received funding from the ARC grant for Concerted Research Actions, financed by the Wallonia-Brussels Federation. TRAPPIST is funded by the Belgian Fund for Scientific Research (Fond National de la Recherche Scientifique, FNRS) under the grant FRFC 2.5.594.09.F, with the participation of the Swiss National Science Fundation (SNF). MG and EJ are F.R.S.-FNRS Senior Research Associate.

This publication benefits from the support of the French Community of Belgium in the context of the FRIA Doctoral Grant awarded to MT.

We acknowledge the use of public TESS data from pipelines at the TESS Science Office and at the TESS Science Processing Operations Center. 
We acknowledge the use of public TESS Alert data from pipelines at the TESS Science Office and at the TESS Science Processing Operations Center. 
Resources supporting this work were provided by the NASA High-End Computing (HEC) Program through the NASA Advanced Supercomputing (NAS) Division at Ames Research Center for the production of the SPOC data products. 
Funding for the TESS mission is provided by NASA's Science Mission Directorate. 
This research has made use of the Exoplanet Follow-up Observation Program website, which is operated by the California Institute of Technology, under contract with the National Aeronautics and Space Administration under the Exoplanet Exploration Program. 
This paper includes data collected by the TESS mission that are publicly available from the Mikulski Archive for Space Telescopes (MAST). 
We thank the TESS GI program G03274 PI, Ryan Cloutier, for proposing the target of this work for 2-minute-cadence observations in Sector 30. 
% The TESS team shall assure that the masses of fifty (50) planets with radii less than 4 REarth are determined. 

This work is based upon observations carried out at the Observatorio Astron\'omico Nacional on the Sierra de San Pedro M\'artir (OAN-SPM), Baja California, M\'exico.

This work makes use of observations from the LCOGT network. Part of the LCOGT telescope time was granted by NOIRLab through the Mid-Scale Innovations Program (MSIP). MSIP is funded by NSF.

This work includes data collected at the Vatican Advanced Technology Telescope (VATT) on Mt. Graham. 
This paper includes data taken on the EDEN telescope network. We acknowledge support from the Earths in Other Solar Systems Project (EOS) and Alien Earths (grant numbers NNX15AD94G and 80NSSC21K0593), sponsored by NASA.

Some of the observations in the paper made use of the High-Resolution Imaging instrument Zorro (Gemini program GS-2020B-LP-105). Zorro was funded by the NASA Exoplanet Exploration Program and built at the NASA Ames Research Center by Steve B. Howell, Nic Scott, Elliott P. Horch, and Emmett Quigley. Zorro was mounted on the Gemini South telescope of the international Gemini Observatory, a program of NSF’s OIR Lab, which is managed by the Association of Universities for Research in Astronomy (AURA) under a cooperative agreement with the National Science Foundation. on behalf of the Gemini partnership: the National Science Foundation (United States), National Research Council (Canada), Agencia Nacional de Investigación y Desarrollo (Chile), Ministerio de Ciencia, Tecnología e Innovación (Argentina), Ministério da Ciência, Tecnologia, Inovações e Comunicações (Brazil), and Korea Astronomy and Space Science Institute (Republic of Korea).

This research has made use of the NASA Exoplanet Archive, which is operated by the California Institute of Technology, under contract with the National Aeronautics and Space Administration under the Exoplanet Exploration Program.

This work made use of the following Python packages: 
\texttt{astropy} \citep{astropy:2013,astropy:2018}, 
\texttt{lightkurve} \citep{2018ascl.soft12013L}, 
\texttt{matplotlib} \citep{Hunter:2007}, 
\texttt{pandas} \citep{mckinney-proc-scipy-2010}, 
\texttt{seaborn} \citep{waskom2020seaborn}, 
\texttt{scipy} \citep{2020SciPy-NMeth} and 
\texttt{numpy} \citep{harris2020array}. 

\end{acknowledgements}

%-----------------------------------------------------------------

\clearpage

\bibliographystyle{aa} % style aa.bst
\bibliography{refs}

%-----------------------------------------------------------------

\begin{appendix}

% \clearpage
% \twocolumn

\section{Transit fit posterior distributions} \label{app:corner}
% \FloatBarrier
% \mbox{}

\begin{figure*}[b!]
\centering
\includegraphics[width=\textwidth, height=\textheight, keepaspectratio]{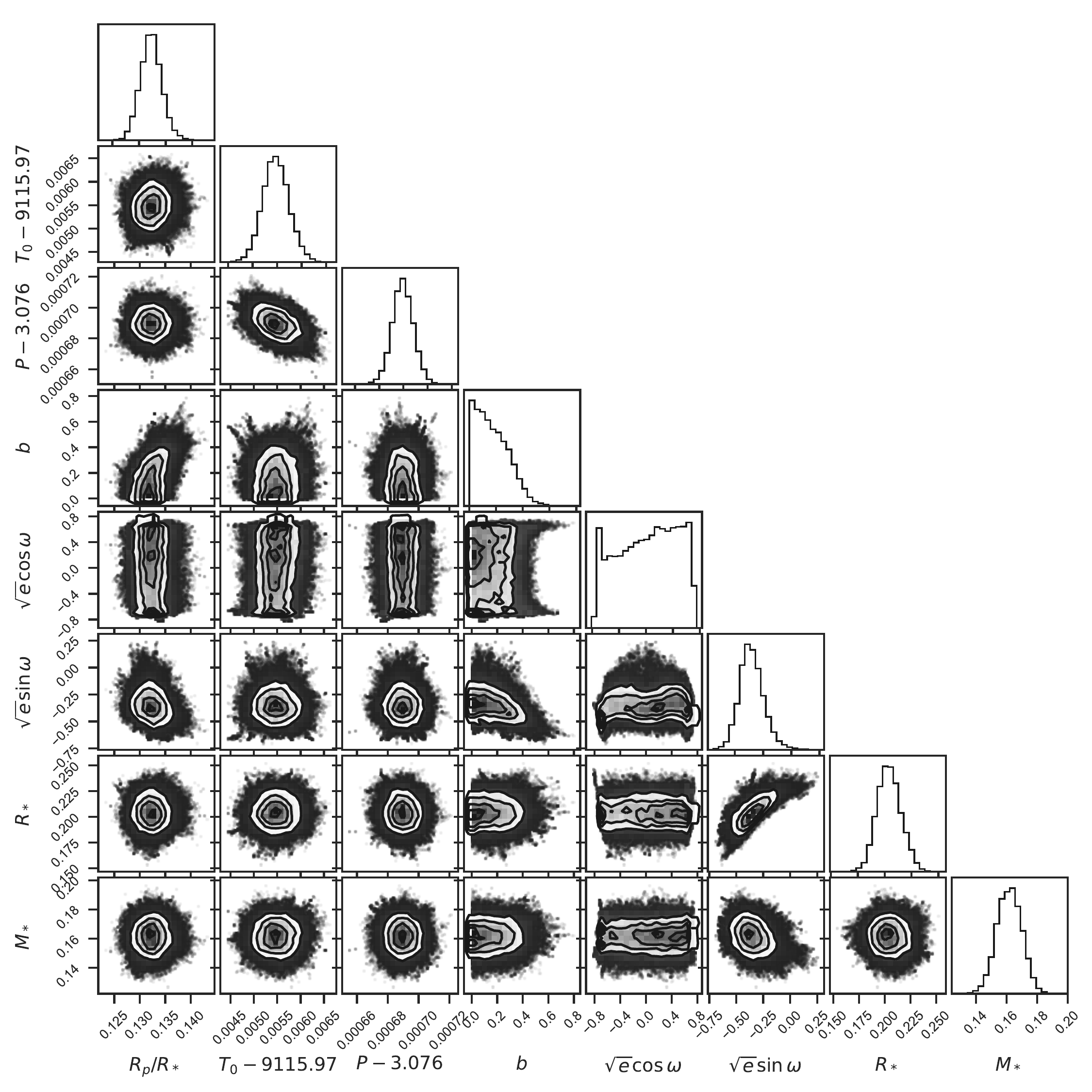}
\caption{MCMC posterior distributions for the fitted transit parameters. 
From left-to-right: the planet-to-star radius ratio, transit epoch (BJD$-$2,450,000), orbital period (d), impact parameter, eccentricity and longitude of periastron parameterised as $\mathrm{\sqrt{e}\cos\omega}$ and $\mathrm{\sqrt{e}\sin\omega}$, and the stellar radius and mass. 
The radius ratio, epoch, period, and stellar parameters can be seen to be Gaussian-like. 
However, the impact parameter and the eccentricity and periastron distributions are non-Gaussian due to their degeneracy with one another. 
The detrending parameters and LDCs, not shown here, are well represented by Gaussian distributions. 
This figure was made using \texttt{corner.py} \citep{corner}.}
\label{fig:corner}
\end{figure*}

% \FloatBarrier
% \vfill
% \afterpage{\clearpage}
\clearpage

\section{Circular orbit fit properties} \label{app:circular}

\begin{table*}[b!]
\caption{Median properties with 1$\sigma$ confidence levels, from the transit analysis. 
We give values for both the eccentric and circular orbit cases. We prefer the use of the eccentric model values in this work. 
We note that the eccentric orbit values given here are identical to those in Table~\ref{table:planet} and are shown as a comparison to the circular orbit values. 
$^\dagger$Calculated with a Bond albedo of zero. 
} 
\label{table:planet-app}
\setlength{\tabcolsep}{12pt}
\centering
\begin{tabular}{l >{\boldmath}c c c}
\hline\hline
Property & {\bf Eccentric orbit} & Circular orbit & Prior \\[0.1cm]
\hline
Fitted parameters: & & & \\[0.1cm]
$\mathrm{T_0}$ (BJD$- 2450000$) & $9115.97547 \pm 0.00027$ & $9115.97501 \pm 0.00027$ & $\mathcal{N}(9115.975, 0.1)$ \\[0.1cm]
P (d) & $3.0766896 \pm 6.5 \times 10^{-6}$ & $3.0766882 \pm 6.5 \times 10^{-6}$ & $\mathcal{N}(3.07665, 0.001)$\\[0.1cm]
$\mathrm{R_p}$/$\mathrm{R_*}$ & $0.1322 \pm 0.0020$ & $0.1319 \pm 0.0018$ & $\mathcal{U}(0.001, 0.4)$\\[0.1cm]
b ($\mathrm{R_*}$) & $0.16_{-0.11}^{+0.15}$ & $0.097_{-0.068}^{+0.098}$ & $\mathcal{U}(0, 1)$\\[0.1cm]
$\mathrm{\sqrt{e}\cos\omega}$ & $0.06_{-0.55}^{+0.45}$ & $-$ & $\mathcal{U}(-1, 1)$ \\[0.1cm]
$\mathrm{\sqrt{e}\sin\omega}$ & $-0.358_{-0.095}^{+0.111}$ & $-$ & $\mathcal{U}(-1, 1)$ \\[0.1cm]
$\mathrm{R_*}$ ($\mathrm{R_\odot}$) & $0.204 \pm 0.011$ & $0.227 \pm 0.005$ & $\mathcal{N}(0.202, 0.011)$ \\[0.1cm]
$\mathrm{M_*}$ ($\mathrm{M_\odot}$) & $0.162 \pm 0.008$ & $0.156 \pm 0.008$ & $\mathcal{N}(0.162, 0.008)$ \\[0.4cm]

Limb-darkening: & & & \\[0.1cm]
$\mathrm{u_1}$ TESS & $0.313 \pm 0.059$ & $0.315 \pm 0.058$ & $\mathcal{N}(0.297, 0.060)$ \\[0.1cm]
$\mathrm{u_2}$ TESS & $0.39 \pm 0.11$ & $0.39 \pm 0.11$ & $\mathcal{N}(0.346, 0.115)$ \\[0.1cm]
$\mathrm{u_1}$ $\mathrm{z^\prime}$ & $0.240 \pm 0.045$ & $0.236 \pm 0.044$ & $\mathcal{N}(0.223, 0.047)$ \\[0.1cm]
$\mathrm{u_2}$ $\mathrm{z^\prime}$ & $0.354 \pm 0.088$ & $0.337 \pm 0.088$ & $\mathcal{N}(0.303, 0.097)$ \\[0.1cm]
$\mathrm{u_1}$ $\mathrm{i^\prime}$ & $0.337 \pm 0.066$ & $0.333 \pm 0.066$ & $\mathcal{N}(0.321, 0.071)$ \\[0.1cm]
$\mathrm{u_2}$ $\mathrm{i^\prime}$ & $0.37 \pm 0.11$ & $0.36 \pm 0.11$ & $\mathcal{N}(0.377, 0.128)$ \\[0.1cm]
$\mathrm{u_1}$ V & $0.56 \pm 0.13$ & $0.54 \pm 0.13$ & $\mathcal{N}(0.615, 0.143)$ \\[0.1cm]
$\mathrm{u_2}$ V & $0.21 \pm 0.17$ & $0.20 \pm 0.17$ & $\mathcal{N}(0.286, 0.182)$ \\[0.1cm]
$\mathrm{u_1}$ Exo & $0.268 \pm 0.064$ & $0.267 \pm 0.065$ & $\mathcal{N}(0.330, 0.069)$ \\[0.1cm]
$\mathrm{u_2}$ Exo & $0.25 \pm 0.12$ & $0.27 \pm 0.12$ & $\mathcal{N}(0.359, 0.124)$ \\[0.4cm]

Derived parameters: & & \\[0.1cm]
$\mathrm{R_p}$ ($\mathrm{R_\oplus}$) & $2.94_{-0.16}^{+0.17}$ & $3.27_{-0.08}^{+0.09}$ &  \\[0.1cm]
$\mathrm{\rho_*}$ & $26.9_{-4.4}^{+4.9}$ & $18.8 \pm 0.9$ &  \\[0.1cm]
$\mathrm{a / R_*}$ & $24.0_{-1.1}^{+1.0}$ & $24.0_{-1.1}^{+1.0}$ &  \\[0.1cm]
a (AU) & $0.0228 \pm 0.0016$ & $0.0254_{-0.0013}^{+0.0012}$ &  \\[0.1cm]
i ($^\circ$) & $89.63_{-0.35}^{+0.27}$ & $89.77_{-0.24}^{+0.16}$ &  \\[0.1cm]
e & $0.26_{-0.12}^{+0.27}$ & $0$ &  \\[0.1cm]
$\mathrm{\omega}$ ($^\circ$) & $279_{-63}^{+47}$ & $-$ &  \\[0.1cm]
$\mathrm{S_p}$ ($\mathrm{S_\oplus}$) & $6.55_{-0.80}^{+0.94}$ & $6.56_{-0.80}^{+0.94}$ &  \\[0.1cm]
$\mathrm{T_{eq}}^\dagger$ (K) & $447 \pm 15$ & $447 \pm 15$ &  \\[0.4cm]

Predicted parameters: & & \\[0.1cm]
$\mathrm{M_p}$ ($\mathrm{M_\oplus}$) & $9.1_{-4.0}^{+7.1}$ & $10.8_{-4.7}^{+8.3}$ &  \\[0.1cm]
K (m/s) & $14.9_{-6.6}^{+12.0}$ & $15.8_{-6.9}^{+12.3}$ &  \\[0.1cm]
TSM & $115_{-50}^{+87}$ & $107_{-47}^{+83}$ &  \\[0.1cm]
ESM & $4.12_{-0.58}^{+0.67}$ & $4.10_{-0.57}^{+0.67}$ &  \\[0.1cm]
\hline
\end{tabular}
\end{table*}

\end{appendix}

\end{document}